\documentclass{article}[12pt]
\usepackage{jcompbio} 

\usepackage{hyperref}       
\usepackage{url}            

\usepackage{graphicx}

\usepackage{amssymb}
\usepackage{amstext}
\usepackage{amsfonts}
\usepackage{amsmath}
\usepackage{amsthm}

\usepackage{algorithm}
\usepackage{algorithmic}

\usepackage{xcolor}

\title{Cotranscriptional kinetic folding of RNA secondary structures including pseudoknots}

\author{Vo Hong Thanh\textsuperscript{1,2}\thanks{thanh.vo@certara.com} 
	\and Dani Korpela\textsuperscript{1}\thanks{dani.korpela@aalto.fi}
	\and Pekka Orponen\textsuperscript{1}\thanks{pekka.orponen@aalto.fi} \\
    \textsuperscript{1}Department of Computer Science, Aalto University  \\
    \textsuperscript{2} Certara, Simcyp Division
}

\newcommand{\highlightred}[1]{\textcolor{black}{#1}}

\begin{document}

\maketitle    

\begin{abstract}
Computational prediction of RNA structures is an important problem in computational structural biology. Studies of RNA structure formation often assume that the process starts from a fully synthesized sequence. Experimental evidence, however, has shown that RNA folds concurrently with its elongation. We investigate RNA secondary structure formation, including pseudoknots, that takes into account the cotranscriptional effects. We propose a single-nucleotide resolution kinetic model of the folding process of RNA molecules, where the polymerase-driven elongation of an RNA strand by a new nucleotide is included as a primitive operation, together with a stochastic simulation method that implements this folding concurrently with the transcriptional synthesis. Numerical case studies show that our cotranscriptional RNA folding model can predict the formation of conformations that are favored in actual biological systems. Our new computational tool can thus provide quantitative predictions and offer useful insights into the kinetics of RNA folding. 
\end{abstract}


\section{Introduction}
Ribonucleic acid (RNA) is a biopolymer constituted of nucleotides with bases adenine (A), cytosine (C), guanine (G) and uracil (U). The synthesis of an RNA molecule from its DNA template is initiated when the corresponding RNA polymerase binds to the DNA promoter region. RNA has been shown to serve diverse functions in a wide range of cellular processes such as regulating gene expression and acting as an enzymatic catalyst~\cite{Collins_2009,Storz_2002}, and has also recently been used as an emerging material for nanotechnology~\cite{Jasinski_2017}. 

Computational prediction of RNA secondary structures given their sequences is often based on the estimation of changes in free energy, which postulates that thermodynamically an RNA strand will fold into a conformation that yields the minimum free energy (MFE) (see e.g.,~\cite{Fallmann_2017} for a review on the topic). The energy of an RNA secondary structure can be modeled as the sum of energies of strand loops flanked by base pairs. The loop energy parameters have been measured experimentally and are detailed in a nearest neighbor parameter database (NNDB)~\cite{Turner_2009}. Methods grounded in the thermodynamic framework, e.g., the Zuker algorithm by~\cite{Zuker_1981} and its extensions~\cite{Zuker_1989,Mathews_1999}, can be used to compute pseudoknot-free MFE secondary structures effectively in a bottom-up manner. Recent attempts to extend the Zuker algorithm to find MFE secondary structures with certain classes of pseudoknots are also proposed~\cite{Rivas_1999,Reeder_2004,Dirks_2003,Akutsu_2000,Chen_2009}; however, finding MFE structures with pseudoknots given a general energy model is a NP-complete problem~\cite{Lyngso_2000}. 

The kinetic approach~\cite{Flamm_2000} is an alternative way to study the RNA folding process. It models the folding as a random process where the additions/deletions of base pairs in the current structure are assigned probabilities proportional to the respective changes in free energy values. A folding pathway of a sequence is then generated by executing stochastic simulation~\cite{Flamm_2000,Mironov_1993,Dykeman_2015}. We refer to~\cite{Marchetti_2017} for a comprehensive review on stochastic simulation and recent work~\cite{Thanh_2014_2,Thanh_2016,Marchetti_2016,Thanh_2017} for state-of-the-art stochastic simulation techniques. Each simulation run on a given RNA sequence can produce a list of possible structures that it can fold into. Such dynamic view of RNA folding allows one to capture cases where local conformations are progressively folded to create metastable structures that kinetically trap the folding, thus complementing the prediction of equilibrium MFE structures produced by the thermodynamic approach. 

The study of RNA structure formation often assumes that the folding process starts from a fully synthesized open strand, the \emph{denatured state}. However, experimental evidence~\cite{Watters_2016,Pan_2006} has shown that RNA starts folding already concurrently with the transcription. The nucleotide transcription speed varies from $200$ nt/sec (nucleotides per second) in phages, to $20$-$80$ nt/sec in bacteria, and $5$-$20$ nt/sec in humans~\cite{Pan_2006}. The RNA dynamics also occur over a wide range of time-scales where base pairing takes about $10^{-3}$ msec; structure formation is about $10$-$100$ msec; and kinetically trapped conformations can persist for minutes or hours~\cite{Hashimi_2008}. One consequence of considering cotranscriptional folding is that the base pairs at the 5' end of the RNA strand will form first, while the ones at the 3' end can only be formed once the transcription is complete, which leads to structural asymmetries. Cotranscriptional folding can thus form \emph{transient} structures that are only present for a specific time period and involved in distinct roles. For instance, gene expression when considering such transient conformations of RNA during cotranscriptional folding can exhibit oscillation behavior~\cite{Bratsun_2005}. We refer to the review by~\cite{Lai_2013} for further discussion on the importance of cotranscriptional effects.

In this work, we extend the kinetic approach to take into account cotranscriptional effects and pseudoknots on the folding of RNA secondary structures at single-nucleotide resolution. Our contribution is twofold. First, we explicitly consider the elongation of RNA during transcription as a primitive action in the model. The time when a new nucleotide is added to the current RNA chain is specified by the transcription speed of the RNA polymerase enzyme. The RNA strand in our modeling approach can elongate with newly synthesized nucleotides added to the sequence and fold simultaneously. To handle the transcription events, we propose an exact stochastic simulation method, the CoStochFold algorithm, to correct the folding pathway. Our method is thus able capture the effects of cotranscriptional folding at single-nucleotide resolution instead of approximating it as in previous approaches~\cite{Flamm_2000,Proctor_2013,Mironov_1986,Zhao_2011,Hua_2018}. Second, our algorithm allows the formation of pseudoknots, which are important for understanding RNA functions. To cope with the challenge in evaluating the energy of pseudoknotted RNA structures, we adapt the NNDB model~\cite{Dirks_2003,Andronescu_2010} to calculate their energy values. It is worth noting that determining a reasonable energy model for RNA structures with pseudoknots is still an open question~\cite{Lyngso_2000,Chen_2009}. However, the advantage of our strategy in comparison with other approaches, e.g., \highlightred{adapting polymer theory in protein folding~\cite{Dill_1999} to evaluate energy of pseudoknots~\cite{Isambert_2000}}, is that in the future when experimental data for pseudoknot parameters are established we can readily apply the simulation without revalidating parameters of the energy model. \highlightred{In addition, we facilitate the computation of energy of RNA structures with pseudoknots by employing the tree representation. We generalize the coarse-grained tree representation of pseudoknot-free RNA structures in literature~\cite{Hofacker_2005} to allow also pseudoknotted motifs.} 

The rest of the paper is organized as follows. Sec.~\ref{sec:2} reviews some background on kinetic folding of RNA. In Sec.~\ref{sec:3}, we present our work to extend the model of RNA folding to incorporate the transcription process and handle the formation of pseudoknots. Sec.~\ref{sec:4} reports our numerical experiments on case studies. Concluding remarks are in Sec.~\ref{sec:5}.

\section{Background on kinetic folding}
\label{sec:2}
Let $S_n$ be a linear sequence of length $n$ of four bases A, C, G, and U in which the 5' end is at position $1$ and the 3' end is at position $n$. A base at position $i$ \highlightred{may form a pair} with a base at $j$, denoted by $(i, j)$, if they form a \emph{Watson-Crick pair} A-U, G-C or a \emph{wobble pair} G-U. A secondary structure formed by intra-molecular interactions between bases in $S_n$ is a list of base pairs $(i, j)$ with $i < j$ satisfying constraints: a) the $i$th base and $j$th base must be separated by at least $3$ (un-paired) bases, i.e., $j - i > 3$; b) for any base pair $(k, l)$ with $k < l$, if $i = k$ then $j = l$; and c) for any base pair $(k, l)$ with $k < l$, if $i < k$ then $i < k < l < j$. The first condition prevents the RNA backbone from bending too sharply. The second one prevents the forming of tertiary structure motifs such as base triplets and G-quartets. The last constraint ensures that no two base pairs intersect, i.e., there are no pseudoknots. We will relax this constraint in Sec.~\ref{sec:3} to allow for the formation of pseudoknots during the folding. 

\begin{figure*}[!htbp]
	\centering
	\includegraphics[scale = 1]{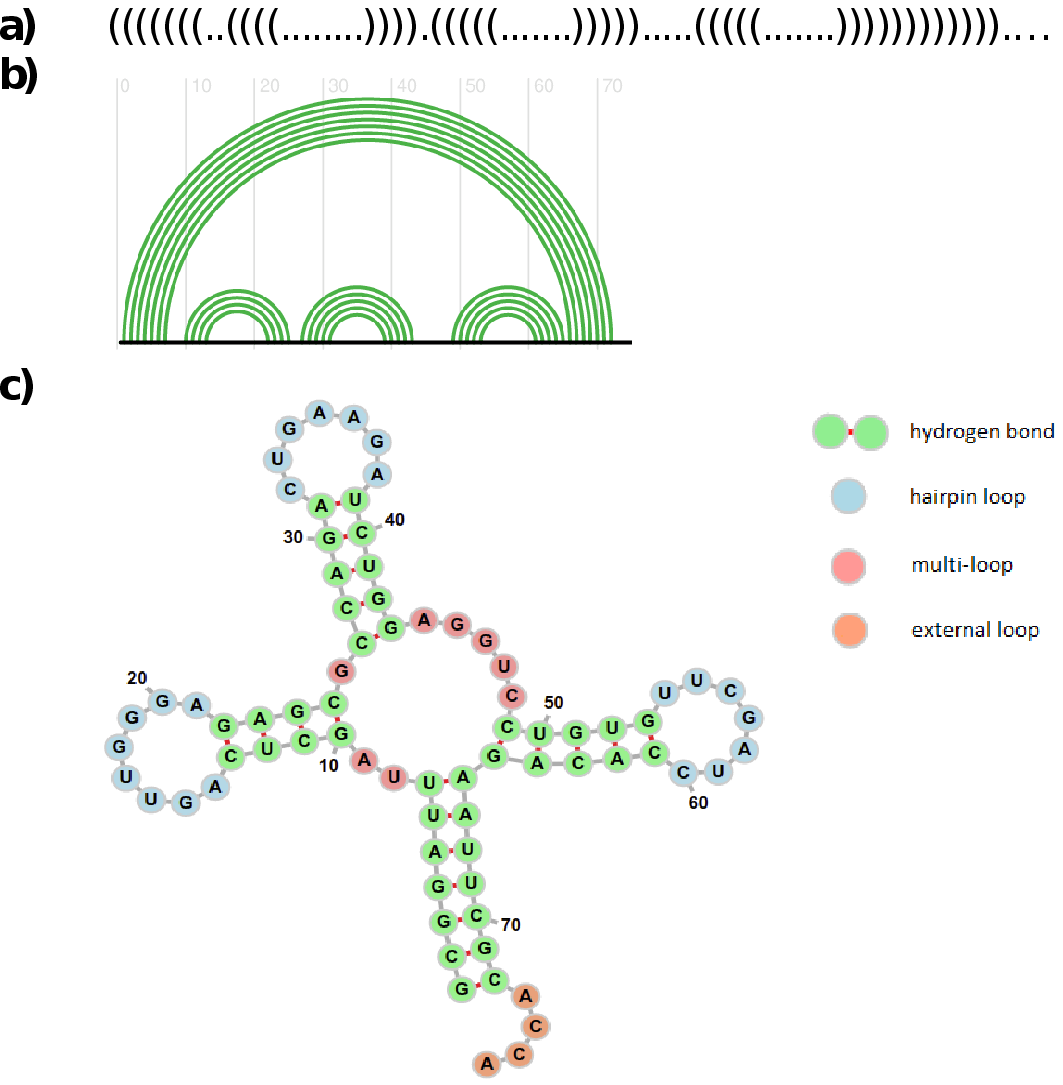}
	\caption[Representation of the tRNA molecule]{Representation of the tRNA molecule in a) dot-bracket notation, b) arc diagram and c) graphical visualization. The graphical visualization is made by the Forna tool~\cite{Kerpedjiev_2015}}.
	\label{fig:rna-representation}       
\end{figure*}

Let $\Omega_{S_n}$ be the set of all possible secondary structures formed by $S_n$. Consider a secondary structure $x \in \Omega_{S_n}$. It can be represented compactly as a string of dots and brackets (see Fig.~\ref{fig:rna-representation}). Specifically, for a base pair $(i, j)$, an opening parenthesis '(' is put at $i$th position and a closing parenthesis ')' at $j$th position. Finally, unpaired positions are represented by dots '.'. The dot-bracket representation is unambiguous because the base pairs in a secondary structure do not cross each other. An alternative method of representing RNA secondary structures is {\em arc diagram}. The arc diagram depicts an RNA structure as a horizontal line from 5' end (left) to 3' end (right) with arcs connecting nucleotides at positions in the sequence to show respective base pairs in the structure. The advantage of the arc diagram is that it can represent RNA structures, e.g., pseudo-knotted structures, that are difficult or impossible to visualize as planar diagrams. Fig.~\ref{fig:rna-representation} a), b) and c), respectively, show the dot bracket, the arc diagram and the graphical visualization of tRNA molecule.  

The free energy of $x$ can be estimated by the \emph{nearest neighbor} model~\cite{Mathews_1999}, in which the free energy of an RNA secondary structure is taken to be the sum of energies of components flanked by base pairs. Formally, for a base pair $(i, j)$ in $x$, we say that base $k$, $i < k < j$, is \emph{accessible} from $(i, j)$ if there is no other base pair $(i', j')$ such that $i < i' < k < j' < j$. The set of accessible bases flanked by base pair $(i, j)$ is called the \emph{loop} $\mathbf{L}(i,j)$. The number of unpaired bases in a loop $\mathbf{L}(i,j)$ is its \emph{size}, while the number of enclosed base pairs determines its \emph{degree}. Based on these properties, loops $\mathbf{L}(i,j)$ can be classified as \emph{stacks} (or \emph{stems}), \emph{hairpins}, \emph{bulges}, \emph{internal loops} and \emph{multi-loops} (or \emph{multi-branch loops}). The unpaired bases that are not contained in loops constitute the \emph{exterior} (or external) loop $\mathbf{L}_e$.

A secondary structure $x$ is thus uniquely decomposed into a collection of loops $x = \cup_{(i, j)} \mathbf{L}(i, j) \cup \mathbf{L}_e$. Based on this decomposition, the free energy $G_x$ (in kcal) of secondary structure $x$ is computed as:
\begin{equation}
G_x = \sum_{(i, j)} G_{\mathbf{L}(i, j)} + G_{\mathbf{L}_e}
\end{equation}
where $G_{\mathbf{L}(i,j)}$ is the free energy of loop $\mathbf{L}(i,j)$. Experimental energy values for $G_{\mathbf{L}(i,j)}$ are available in the nearest neighbor database~\cite{Turner_2009}. 

Let $y \in \Omega_{S_n}$ be a secondary structure derived directly from $x$ by an intramolecular reaction between bases $i$ and $j$ in $x$. Commonly, three operations on a pair of bases, referred to as the \emph{move set} (see Fig.~\ref{fig:operations-rna}), are defined~\cite{Flamm_2000}:
\begin{itemize}
	\item \emph{Addition:} $y$ is derived by adding a base pair that joins bases $i$ and $j$ in $x$ that are currently unpaired and eligible to pair.
	\item \emph{Deletion:} $y$ is derived by breaking a current base pair $(i, j)$ in $x$.
	\item \emph{Shifting:} $y$ is derived by shifting a base pair $(i, j)$ in $x$ to form a new base pair $(i, k)$ or $(k, j)$.
\end{itemize}
Let $k_{x \rightarrow y}$ be the rate (probability per time unit) of the transition from $x$ to $y$. In a conformation $x$, the RNA molecule may wander vibrationally around its energy basin for a long time, before it surmounts an energy barrier to escape to a conformation $y$ in another basin. The dynamics of the transition from $x$ to $y$ characterizes a rare event in Molecular Dynamics (MD). Here, we adopt the coarse-grained kinetic Monte Carlo approximation~\cite{Metropolis_1953,Kawasaki_1966}, and model the transition rate $k_{x \rightarrow y}$ as: 
\begin{equation}
\label{rate}
k_{x \rightarrow y} = k_0 e^{{-\Delta G_{xy}}/{2RT} }
\end{equation}	
where $T$ is absolute temperature in Kelvin (K), $R = 1.98717 \times 10^{-3} (kcal \cdot K^{-1} \cdot mol^{-1})$ is the gas constant and $\Delta G_{xy} = G_y - G_x$ denotes the difference between free energies of $x$ and $y$. The constant $k_0$, normally taking values in the range $10^{-4}$ to $10^{-3}$, provides a calibration of time. 

Let $P(x, t)$ be the probability that the system is at conformation $x$ at time $t$. The dynamics of $P(x, t)$ is formulated by the (chemical) master equation~\cite{Marchetti_2017} as:
\begin{equation}
\label{cme}
\frac {dP(x, t)} {dt} = \sum_{y \in \Omega_{S_n}} \Big[k_{y \rightarrow x}P(x, t) - k_{x \rightarrow y} P(x, t) \Big]
\end{equation}
Analytically solving Eq.~\ref{cme} requires to enumerate all possible states $x$ and their neighbors $y$. The size of the state space $\|\Omega_{S_n}\| \sim n^{-3/2} \alpha^n$ with $\alpha = 1.8488$ increases exponentially with the sequence length $n$, and the number of neighbors of $x$ is in order of $O(n^2)$~\cite{Hofacker_1998}. Thus, due to the high dimension of the state space, solving Eq.~\ref{cme} often involves numerical simulation. 

Let $P(y, \tau|x, t)$ be the probability that, given current structure $x$ at time $t$, $x$ will fold into $y$ in the next infinitesimal time interval $[t + \tau, t + \tau + d\tau)$. We have 
\begin{equation}
\label{pdf}
P(y, \tau|x, t) = k_{x \rightarrow y} e^{-k_x \tau}d\tau
\end{equation}	 
where $k_x = \sum_{y \in \Omega_{S_n}}{k_{x \rightarrow y}}$ is the sum of transition rates to single-move neighbors of $x$. \highlightred{Eq.~\ref{pdf} lays down the mathematical framework for stochastic RNA folding. Integrating Eq.~\ref{pdf} with respect to $\tau$ from $0$ to $\infty$, the probability that $x$ moves to $y$ is $k_{x \rightarrow y} / k_{x}$. Summing Eq.~\ref{pdf} over all possible states $y \in \Omega_{S_n}$, it shows the waiting time $\tau$ until the transition occurs follows an exponential distribution $Exp(k_{x})$}. These facts are the basis for our kinetic folding algorithm called StochFold presented as Algorithm~\ref{alg:stochfold}. We note that StochFold shares the structure of the earlier algorithm Kinfold~\cite{Flamm_2000} and its improvements~\cite{Dykeman_2015,Thanh_2014}.

\begin{algorithm}
	\caption{StochFold}
	\label{alg:stochfold}
	\begin{algorithmic}[1]	
		\REQUIRE initial RNA conformation $s_0$ and ending time $T_{max}$					
		\STATE initialize $x = s_0$ and time $t = 0$ 
		\REPEAT
		\STATE enumerate next possible conformations of the current conformation $x$ and put into set $Q$ 
		\STATE compute the transition rate $k_{x \rightarrow y} $ for each $y \in Q$ and total rate $k_x = \sum_{y \in Q}{k_{x \rightarrow y}}$
		\STATE select next conformation $y \in Q$ with probability $k_{x \rightarrow y} / k_x$
		\STATE sample waiting time to the next folding event $\tau \sim Exp(k_x)$
		\STATE set $x = y$ and $t = t + \tau$
		\UNTIL{$t \geq T_{max}$}	
	\end{algorithmic}
\end{algorithm}

\section{Cotranscriptional kinetic folding of RNA}
\label{sec:3}
\begin{figure}[!htbp]
	\centering
	\includegraphics[scale = 0.5]{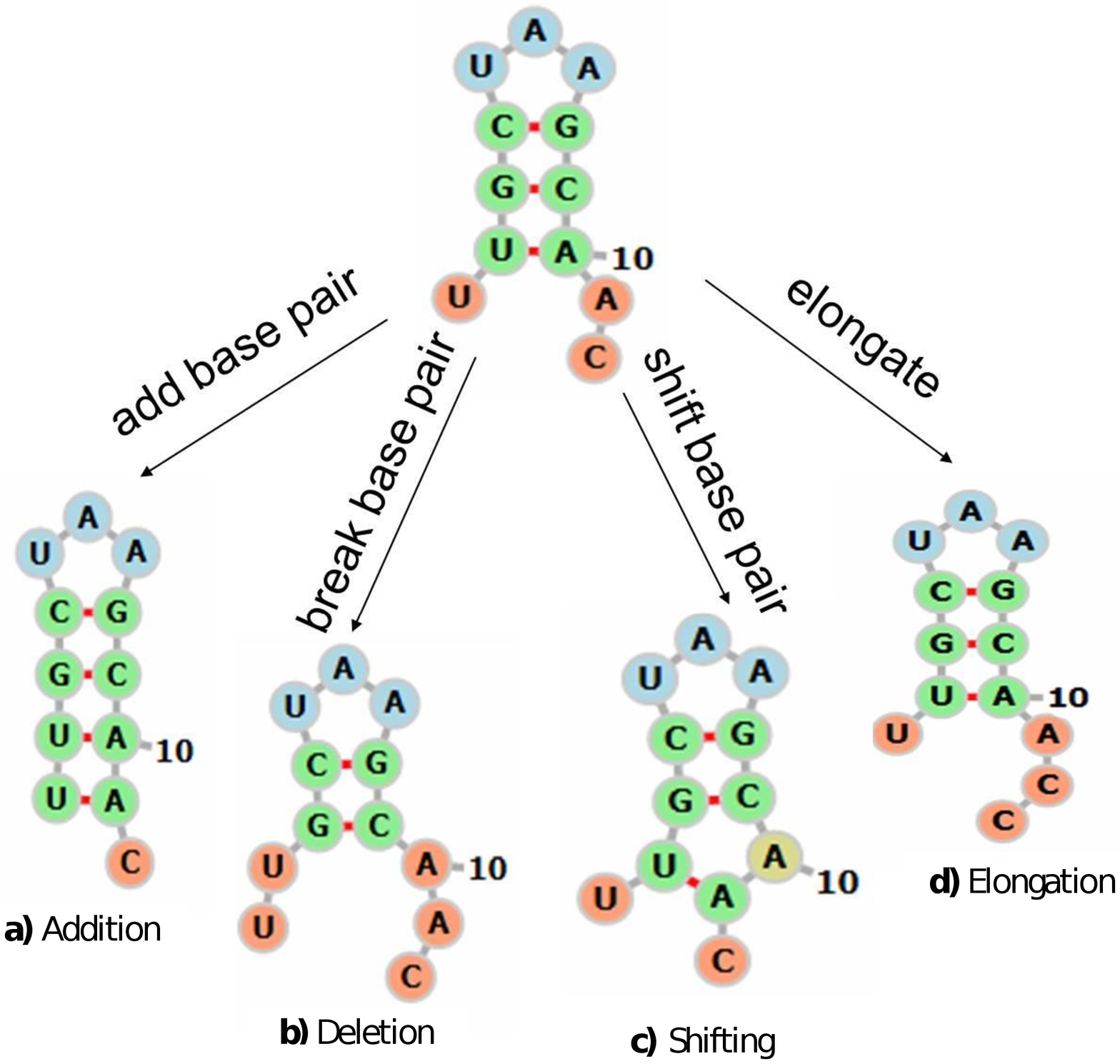}
	\caption[Extended move set]{Extended move set consisting of a) addition, b) deletion, c) shifting and d) elongation. The elongation move models the transcription process extending the current RNA chain with a new nucleotide at the 3' end.}
	\label{fig:operations-rna}       
\end{figure}

The folding of an RNA strand adapts immediately to new nucleotides synthesized during the transcription. The kinetic approach described in Sec.~\ref{sec:2} cannot capture the effects of such cotranscriptional folding, because it considers only interactions between bases already present in the sequence. We outline in this section an approach to incorporating these effects in the simulation. The transcription process is explicitly taken into account by extending the move set with the new operation of \textit{elongation}. Our extended move set thus comprises four operations: addition, deletion, shifting and elongation. The first three operations are defined as in the previous section. In elongation, the current RNA chain increases in length and a newly synthesized nucleotide is added to its 3' end. Figure~\ref{fig:operations-rna} illustrates the extended move set.

Under the extended move set, we define two event types: \emph{folding event} and \emph{transcription event}. A folding event is an \emph{internal event} that occurs when one of the three operations addition, deletion or shifting, is applied to a base pair of the current sequence. A transcription event happens when the elongation operation is applied. It is an \emph{external event} whose rate is specified by the transcription speed of the RNA polymerase enzyme. The occurrences of transcription events break the Markovian property of transitions between conformations. This is because when a new nucleotide is added to the current RNA conformation, the number of next possible conformations increases. The waiting time of the next folding event also changes and thus a new folding event has to be recomputed.  

Algorithm~\ref{alg:costochfold} outlines how the CoStochFold algorithm handles this situation. The key element of CoStochFold (lines~\ref{begin-search} - \ref{end-search}) is a race where the event having the smallest waiting time will be selected to update the current RNA conformation. More specifically, suppose the current structure is $x$ at time $t$. Let $\tau_e$ be the waiting time to the next folding event and $\tau_{trans}$ the waiting time to the next transcription event. Assuming that no events occur earlier, $\tau_e$ has an exponential distribution with rate $k_x$ which is the sum of all transition rates of applying addition, deletion and shifting operations to base pairs in $x$. For simplifying the computation of $\tau_{trans}$, we assume that it is the expected time to transcribe one nucleotide. Let $N_{trans}$ be the (average) transcription speed of the polymerase. We compute $\tau_{trans}$ as: 
\begin{equation}
\tau_{trans} = 1 / N_{trans} 
\end{equation}
Thus, given current time $t$, the next folding event will occur at time $t_e = t + \tau_e$ and, respectively, the transcription event where a new nucleotide will be added to the current sequence is scheduled at time $t_{trans} = t + \tau_{trans}$. We decide which event will occur by comparing $t_e$ and $t_{trans}$. If $t_e > t_{trans}$, then a new nucleotide is first transcribed and added to the current RNA conformation. Otherwise, a folding event is performed where a structure in the set $Q$ of neighboring structures is selected to update the current conformation.

\begin{algorithm}
	\caption{CoStochFold}
	\label{alg:costochfold}
	\begin{algorithmic}[1]	
		\REQUIRE initial RNA conformation $s_0$, transcription speed $N_{trans}$, and ending time $T_{max}$					
		\STATE initialize $x = s_0$ and time $t = 0$ 
		\STATE set $\tau_{trans} = 1 / N_{trans}$
		\STATE compute the next transcription event $t_{trans} = t + \tau_{trans}$
		\label{transcription-event-time}
		\REPEAT
		\STATE enumerate next possible conformations by applying addition, deletion and shifting operations on the current conformation $x$ and put into set $Q$ 
		\STATE compute the transition rate $k_{x \rightarrow y}$, for $y \in Q$, and total rate $k_x = \sum_{y \in Q}{k_{x \rightarrow y}}$ 
		\STATE sample waiting time to the next folding event $\tau_e \sim Exp(k_x)$ and set $t_e = t + \tau_e$
		
		\IF{($t_e > t_{trans}$)} \label{begin-search}
		\STATE elongate $x$
		\STATE set $t = t_{trans}$
		\STATE compute the next transcription event $t_{trans} = t + \tau_{trans}$		
		\ELSE
		\STATE select next conformation $y \in Q$ with probability $k_{x \rightarrow y}/ k_x$
		\STATE set $x = y$ and $t = t_e$
		\ENDIF	\label{end-search}	
		\UNTIL{$t \geq T_{max}$}	
	\end{algorithmic}
\end{algorithm}

We remark that one can easily extend Algorithm~\ref{alg:costochfold} to allow modeling $\tau_{trans}$ as a random variable without changing the steps of event selection. Specifically, one only needs to change step \ref{transcription-event-time} in  Algorithm~\ref{alg:costochfold} to generate the waiting time of the next transcription event, while keeping the simulation otherwise unchanged.  

\subsection{Handling pseudoknots}
\begin{figure}[!htbp]
	\centering
	\includegraphics[scale = 0.8]{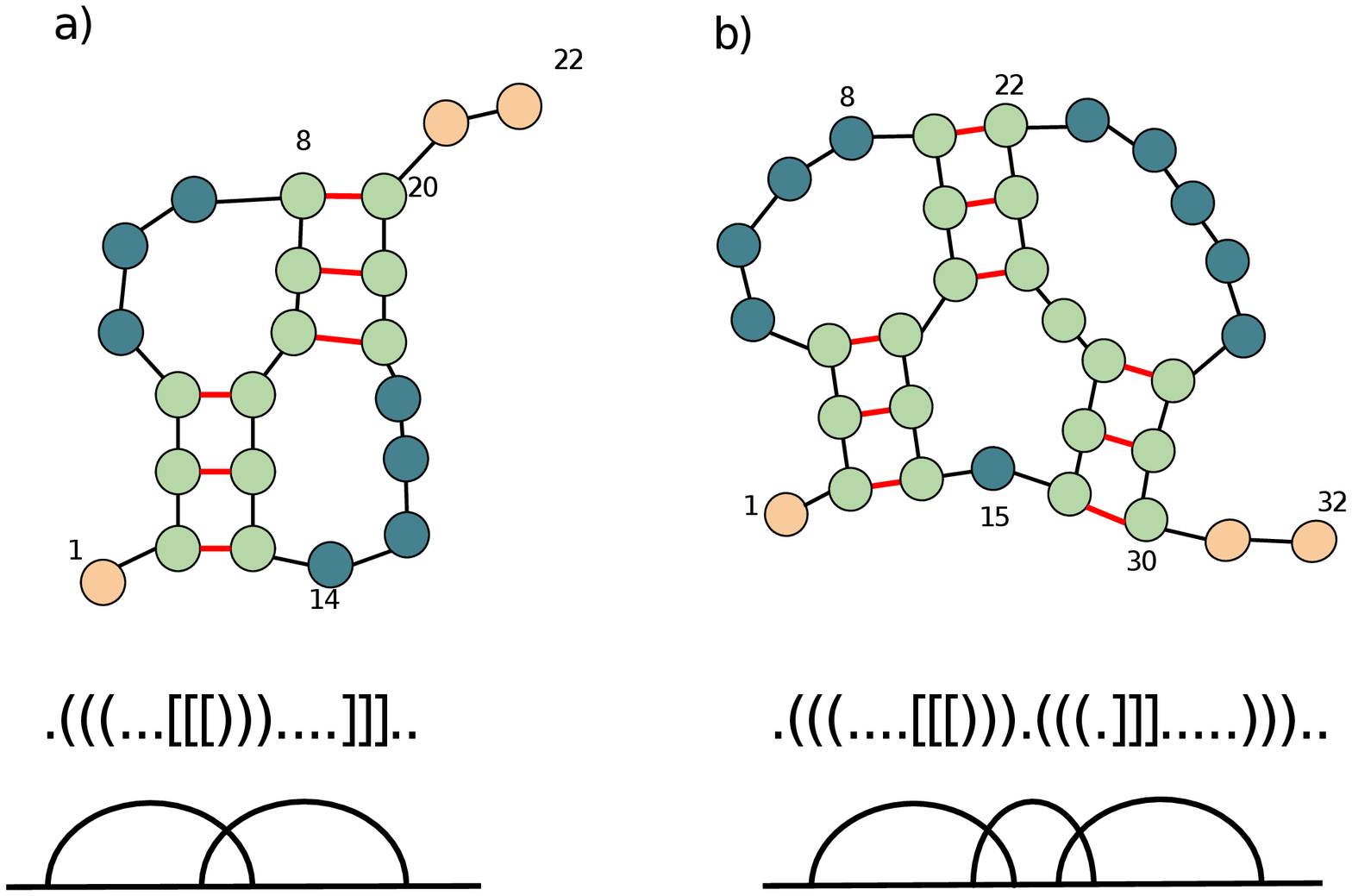}
	\caption[Pseudoknot types]{a) H-type pseudoknot and b) K-type pseudoknot depicted with extended dot-bracket notation and arc diagrams.}
	\label{fig:pseudoknots}       
\end{figure}

This section extends the CoStochFold algorithm to include structures with pseudoknots during the enumeration of neighbor structures (see step 5, Algorithm 2). A pseudoknot occurs if there exists a crossing between two base pairs. Here we restrict to the two most common pseudoknots: the H-type and K-type (kissing hairpin)~\cite{Reidys_2011}. We use the extended dot-bracket notation, i.e., augment the original dot-bracket with additional types of bracket pairs, e.g., [], $\{ \}$ and $\langle \rangle$, to denote the crossing base pairs. Fig.~\ref{fig:pseudoknots} depicts examples of RNA structures with H-type and K-type pseudoknots and their corresponding extended dot-bracket notations and arc diagrams.

Let $\mathbf{L}(i, j)$ be a pseudoknot flanked by the bases $i$ and $j$. \highlightred{We compute its energy $G_{\mathbf{L}(i, j)}$ by adapting the NNDB model~\cite{Dirks_2003,Andronescu_2010,Reidys_2011}. The energy of a pseudoknot consists of an initiation penalty and structural penalties. The initiation penalty depends on whether the pseudoknot is unnested or nested within another multiloop or pseudoknot. The structural penalty takes into account the number of unpaired bases, nested substructures and the energy of the pseudoknotted stems.} Specifically, the energy of $\mathbf{L}(i, j)$ is calculated by the formula:
\begin{equation}
 \label{pseudoE}
	 G_{\mathbf{L}(i, j)} = \beta_{\mathbf{L}(i, j)} + P*\beta_2 + U*\beta_3
\end{equation}
where $\beta_{\mathbf{L}(i, j)}$ is an initiation energy term that penalizes the formation of the pseudoknot, and $P$ and $U$, respectively, denote the numbers of paired bases that flank the interior of the pseudoknot and unpaired bases inside the pseudoknot. The corresponding parameters $\beta_2$ and $\beta_3$ are used to penalize the formation of base pairs $P$ and unpaired bases $U$ correspondingly.

\begin{figure}[!htbp]
	\centering
	\includegraphics[scale = 0.68]{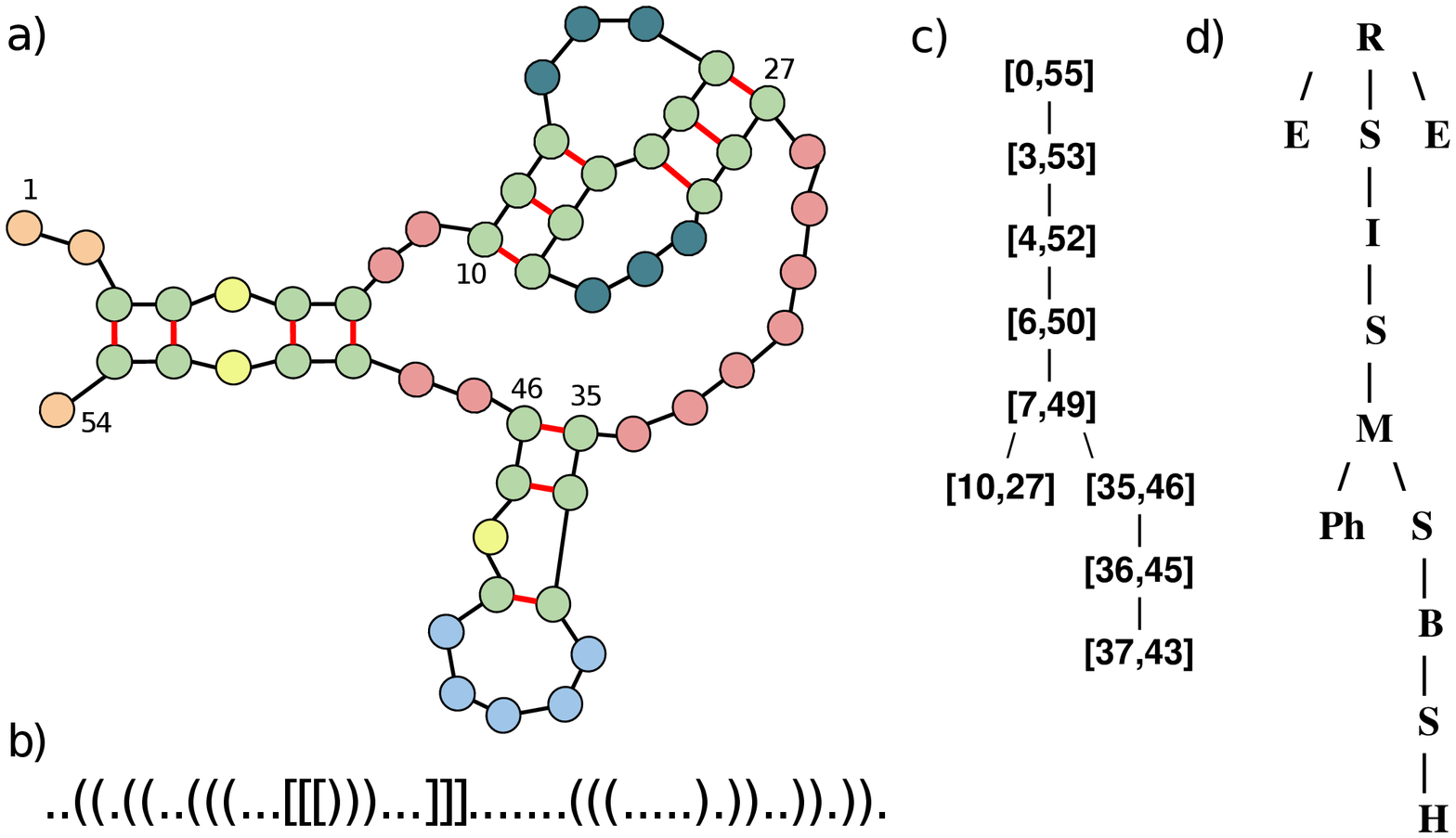}
	\caption[An example of a motif tree]{An example of a motif tree. a) Secondary structure with pseudoknot, b) its extended dot-bracket form, c) closed region tree, and d) motif tree. Starting from the root R (a dummy node) the motif tree represents the relationship of loops: exterior (E), stem (S), hairpin (H), multi-branch (M), pseudoknot (Ph), bulge (B) in the structure.}
	\label{fig:MotifTree}       
\end{figure}

To facilitate the evaluation of the energy of an RNA structure $x$ with pseudoknots, we first parse $x$ to {\em closed regions}~\cite{Condon_2007}. A set of bases $\{i, i + 1,...,j\}$ is called a closed region if a) no base in the region pairs to a base outside of the interval $\{i, i + 1,...,j\}$, and b) such region cannot be partitioned into smaller closed regions. We then decompose each closed region into loops and pseudoknots. Such structural motifs will form a tree that we called a {\em motif tree}. An example of a motif tree is depicted in Fig.~\ref{fig:MotifTree}. Having the motif tree for structure $x$, we can traverse it from the leaves to the root to obtain its energy value. Specifically, we evaluate energy values of motifs at the leaves and send them to their parents. At each inner node, we sum of its energy and the child nodes, then propagate to the upper level. The process is done recursively until reaching the root where total energy sum $G_x$ is returned.   
       
\section{Numerical experiments}
\label{sec:4}
We illustrate the application of our cotranscriptional kinetic folding method on four case studies: a) the E. coli signal recognition particle (SRP) RNA~\cite{Watters_2016}, b) the switching molecule~\cite{Flamm_2000}, c) the Beet soil-borne virus~\cite{pseudobase++} and d) the SV-11 variant in Q$\beta$ replicase~\cite{Biebricher_1992}. We use these examples to manifest the characteristics of our method that thermodynamic/kinetic methods~\cite{Zuker_1981,Gultyaev_1995,Flamm_2000} would fail to capture if initiated from fully denatured sequences. Our cotranscriptional folding method is not only able to produce these structures, but also provides insight into mechanisms that biological systems may use to guide the structure formation process. \highlightred{Finally, we assess the computational performance of the proposed simulation algorithm on sequences of varying lengths.} The code for the implementation of our CoStochFold algorithm is available at: \fontsize{9}{12} {\url{https://github.com/vo-hong-thanh/stochfold}}. 

\subsection{Signal recognition particle (SRP) RNA}
\begin{figure*}[!htbp]
	\centering
	\includegraphics[scale = 0.75]{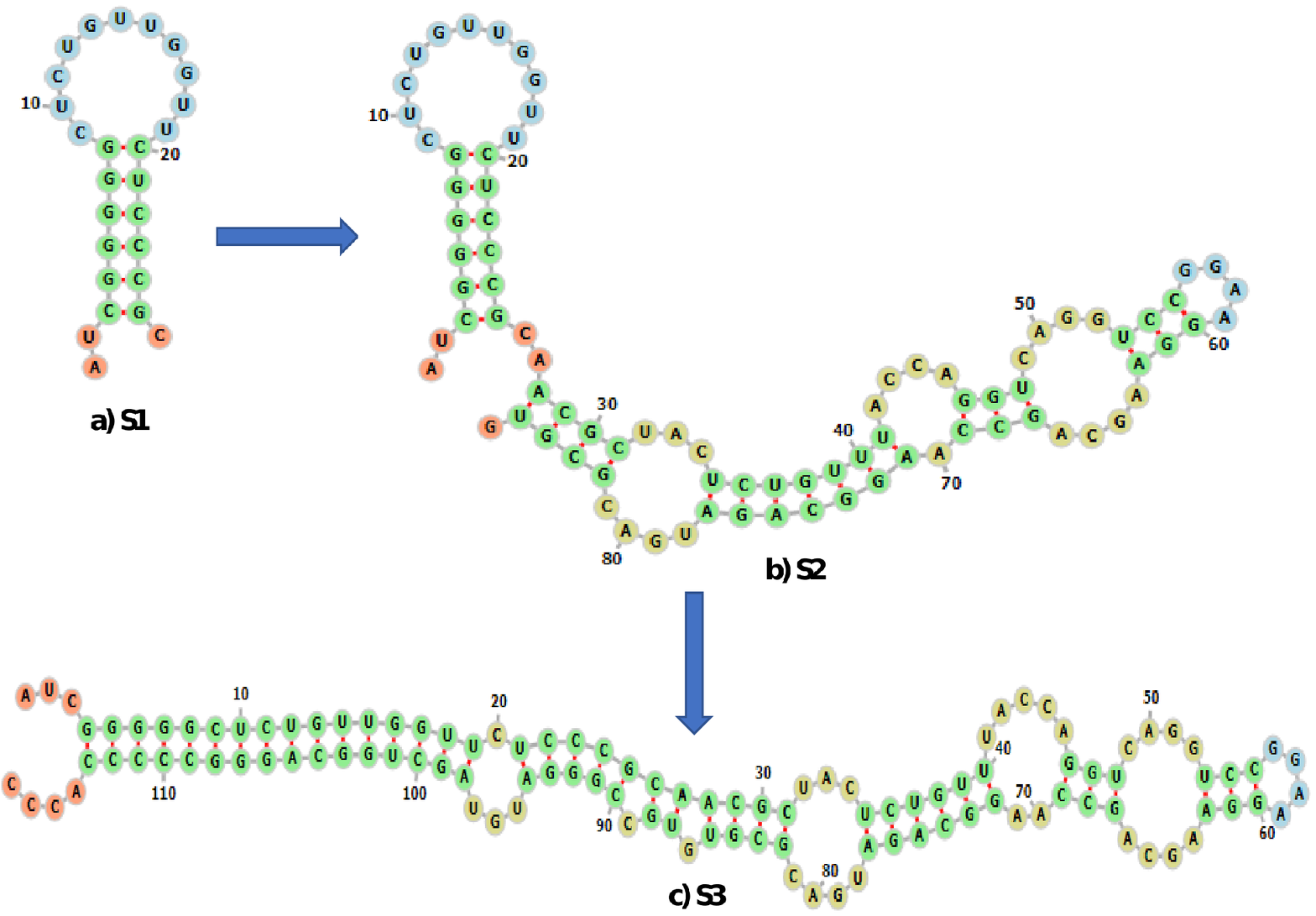}
	\caption[The folding pathway of secondary structures of SRP RNA]{The folding pathway of secondary structures of the E. coli signaling recognition particle (SRP) RNA. The hairpin motif S1 is formed at transcript length $25$nt and form S2 completed at length $86$nt. When reaching transcript length $117$nt, SRP rearranges into its stable helical shape S3. The visualization of structures is made by the Forna tool~\cite{Kerpedjiev_2015}.}
	\label{fig:structural-formation_SRP-1}       
\end{figure*}

This section studies the process of structural formation of the E. coli SRP RNA during transcription. SRP is a $117$nt long molecule, which recognizes the signal peptide and binds to the ribosome locking the protein synthesis. Its active structure is a long helical structure containing interspersed inner loops (see S3 in Fig.~\ref{fig:structural-formation_SRP-1}). Experimental work~\cite{Watters_2016} using SHAPE-seq techniques has suggested a series of structural rearrangements during transcription that ultimately result in the SRP helical structure. In particular, the 5' end of SRP forms a hairpin structure during early transcription. The structure persists until the transcript reaches a length of $117$nt. The unstable hairpin then rearranges to its active structure. Fig.~\ref{fig:structural-formation_SRP-1} depicts three structural motifs at $25$nt (S1), $86$nt (S2), and $117$nt (S3), respectively, in the formation of SRP. Specifically, the hairpin motif S1 emerges at transcript length $25$nt, and the transcript then continues elongating to form structure S2 at length $86$nt. When reaching transcript length $117$nt, SRP rearranges into its persistent helical conformation S3.

\begin{figure*}[!htbp]
	\centering
	\includegraphics[scale = 0.9]{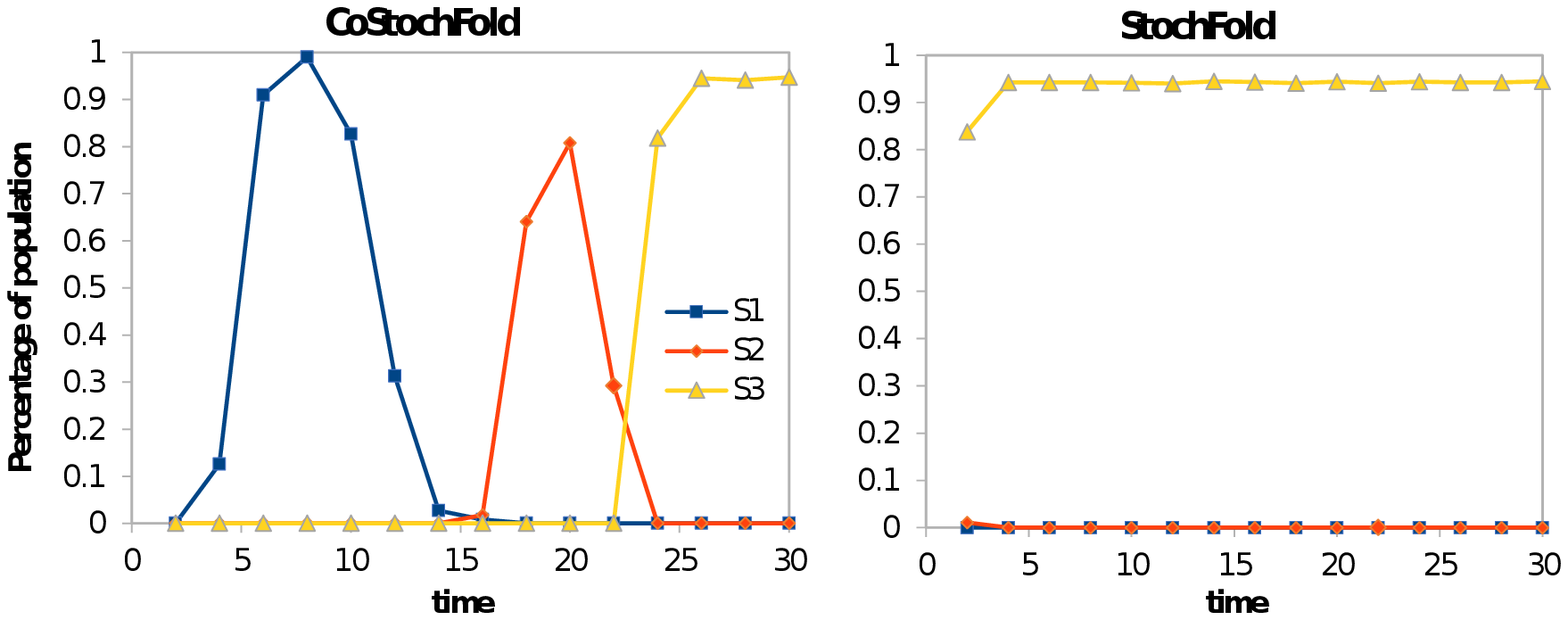}
	\caption[Prediction of the structural formation of SRP]{Prediction of the structural formation of SRP. Left: cotranscriptional folding. Right: folding from denatured state without transcription. The frequency of occurrence of a motif on y-axis is computed as the numbers of occurrences over total $10000$ simulation runs. Time on the x-axis is in seconds of simulated time.}	
	\label{fig:structural-formation_SRP-2}       
\end{figure*} 

We validated the prediction of the CoStochFold algorithm against the experimental work in~\cite{Watters_2016}. To do that, we performed $10000$ simulation runs of the algorithm to fold SRP cotranscriptionally. The average transcription speed was set to $5$ nt/sec. Fig.~\ref{fig:structural-formation_SRP-2} shows the frequency of occurrences of the considered structures during the simulated time of $30$ seconds. Kinetic folding starting from the denatured state was carried out by the StochFold algorithm, while cotranscriptional folding was conducted by the CoStochFold algorithm. The plot on the left shows the cotranscriptional folding of SRP and the plot on the right presents the folding of SRP starting from the denatured state. The figures clearly show that the CoStochFold algorithm can capture the folding pathway of SRP. Specifically, the hairpin motif S1 starts to form at about $t = 4$s when the transcript length is $20$nt and peaks at about $t = 8$s when $40$nt have been transcribed. At about $t = 18$s, Structure S2 appears and then rearranges to S3 at about $t = 24$s. We note that in the simulated folding without considering transcription only the conformation S3 is encountered. 

\subsection{Switching molecule}
We consider the dynamic folding of an artificial RNA sequence $S =$ "GGCCCCUUUGGGGGCCAGACCCCUAAAGGGGUC"~\cite{Flamm_2000}. Two stable conformations of the sequence are: the MFE structure $x =$ ``((((((((((((((.....))))))))))))))'' ($-26.20$ kcal), and a suboptimal structure $y =$ ``((((((....)))))).((((((....))))))'' ($-25.30$ kcal). We use this example to demonstrate how by tuning the transcription speed we can change the ratio of occurrences of structures $x$ and $y$. Here we focus on the number of first-hitting time occurrences of a target structure. The number of first-hitting time occurrences of a structure in a time interval divided by the total number of simulation runs approximates the first-passage time probability of the structure, i.e., its folding time~\cite{Flamm_2000}. 

\begin{figure*}[!htbp]
	\centering
	\includegraphics[scale = 0.85]{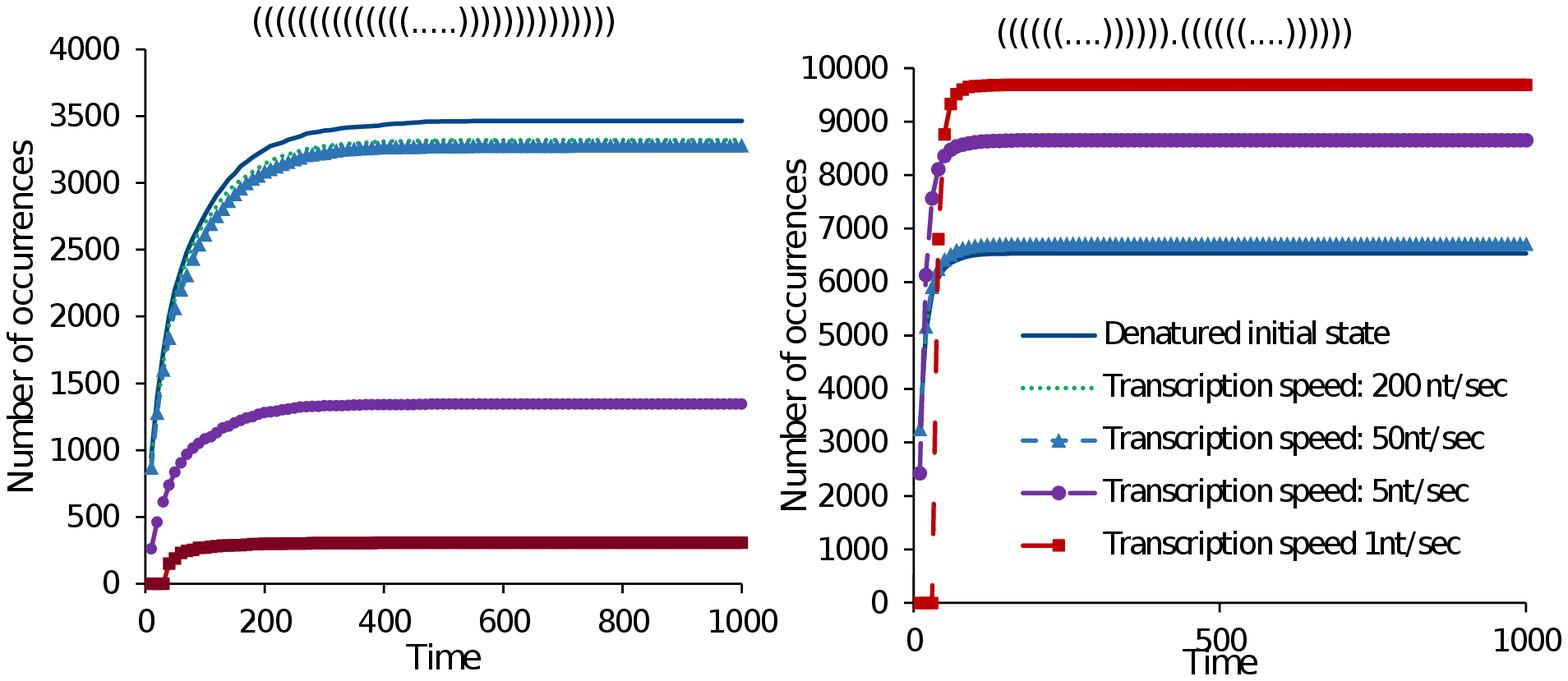}
	\caption[Cumulative first-hitting time occurrences structures]{Cumulative first-hitting time occurrences of MFE structure $x =$ ``((((((((((((((.....))))))))))))))'' ($-26.20$ kcal, left) and suboptimal $y =$ ``((((((....)))))).((((((....))))))'' ($-25.30$ kcal, right). Time on the x-axis is in seconds of simulated time.}
	\label{fig:switching-molecule-1}       
\end{figure*}

\begin{figure}[!htbp]
	\centering
	\includegraphics[scale = 0.8]{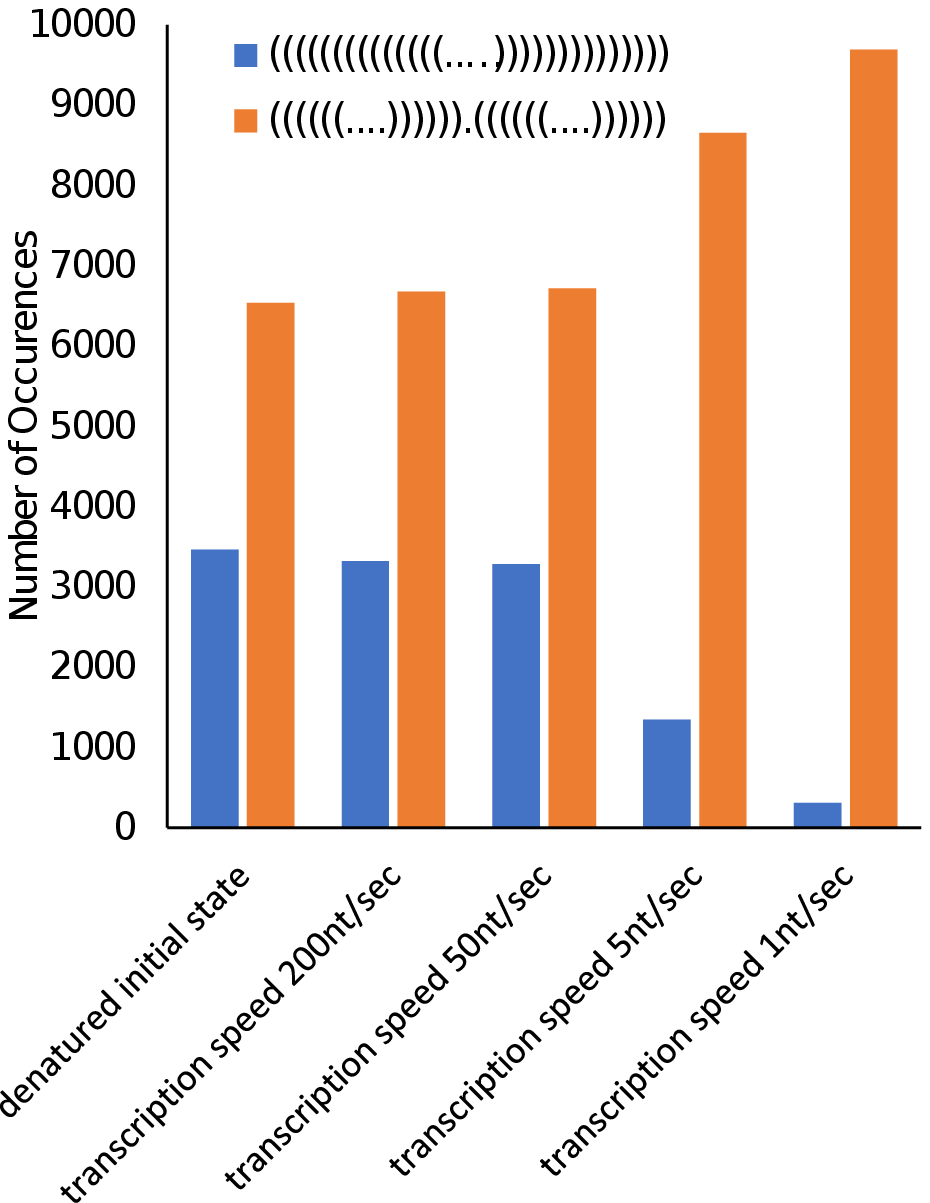}
	\caption[Total number of occurrences of structures by varying transcription speeds]{Total number of occurrences of MFE structure $x =$ "((((((((((((((.....))))))))))))))" ($-26.20$ kcal) and suboptimal $y =$ "((((((....)))))).((((((....))))))" ($-25.30$ kcal) with simulated time $T_{max} = 1000$ seconds by varying transcription speeds.}
	\label{fig:switching-molecule-2}       
\end{figure}

Fig.~\ref{fig:switching-molecule-1} plots the number of first-hitting time occurrences of the MFE structure $x$ and the suboptimal $y$ with varying transcription speeds. We performed $10000$ simulation runs of the CoStochFold algorithm on the sequence $S$ in which each simulation ran until a target structure was observed or the ending time $T_{max} = 1000$ seconds was reached. The constant $k_0 = 1$ in Eq.~\ref{rate} is used in this case study to scale the time. Fig.~\ref{fig:switching-molecule-1} shows that changing the transcription speed of the polymerase significantly affects the folding characteristics of the sequence. Specifically, cotranscriptional folding with slow transcription speed favors the suboptimal structure $y$. It increases the number of occurrences of $y$, while reducing the number of occurrences of the MFE structure $x$.

Fig.~\ref{fig:switching-molecule-2} compares the total number of first-hitting time occurrences of the MFE structure $x$ with respect to the suboptimal conformation $y$ up to time $T_{max} = 1000$. We note that if the simulation starts from the fully denatured state, the occurrence ratio of the suboptimal conformation $y$ to the MFE structure $x$ is about 2:1, as also observed by~\cite{Flamm_2000}. However, the ratio increases noticeably when the transcription speed decreases. For example, the occurrence ratio of the suboptimal conformation $y$ to the MFE structure $x$ is about 6.5:1 in the case of transcription speed $5$ nt/sec.

\subsection{Beet soil-borne virus}
\label{sec:4.4}
\begin{figure}[!htbp]
	\centering
	\includegraphics[scale = 0.5]{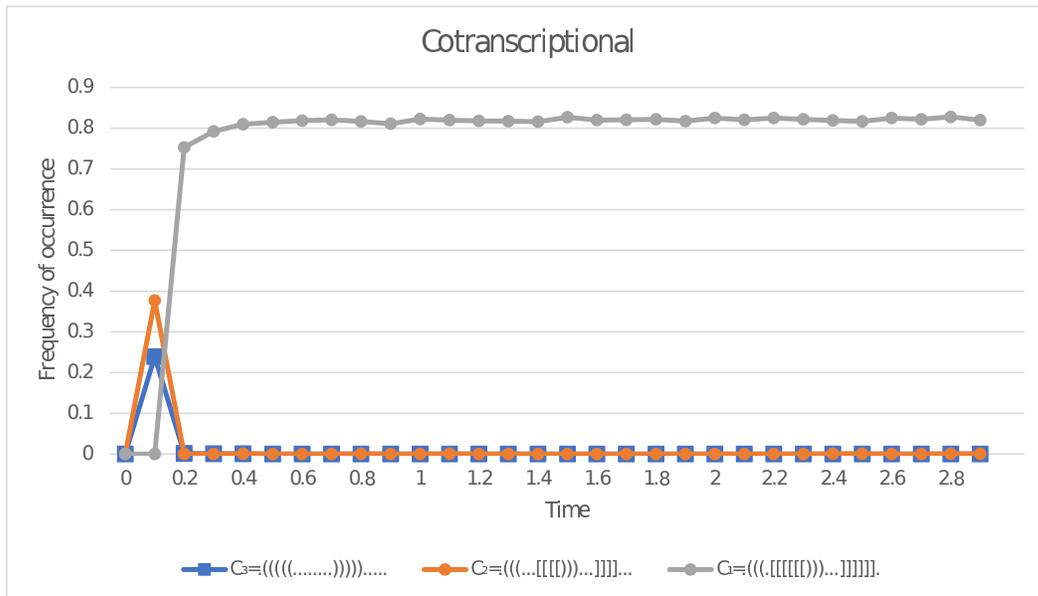}
	\caption[Cotranscriptional folding of the Beet soil-borne virus]{ \highlightred{Cotranscriptional folding of the Beet soil-borne virus with the most frequent and two intermediate structures. Time on the x-axis is in seconds of simulated time.}}
	\label{fig:CotransPseudo}       
\end{figure}

\begin{figure}[!htbp]
	\centering
	\includegraphics[scale = 0.5]{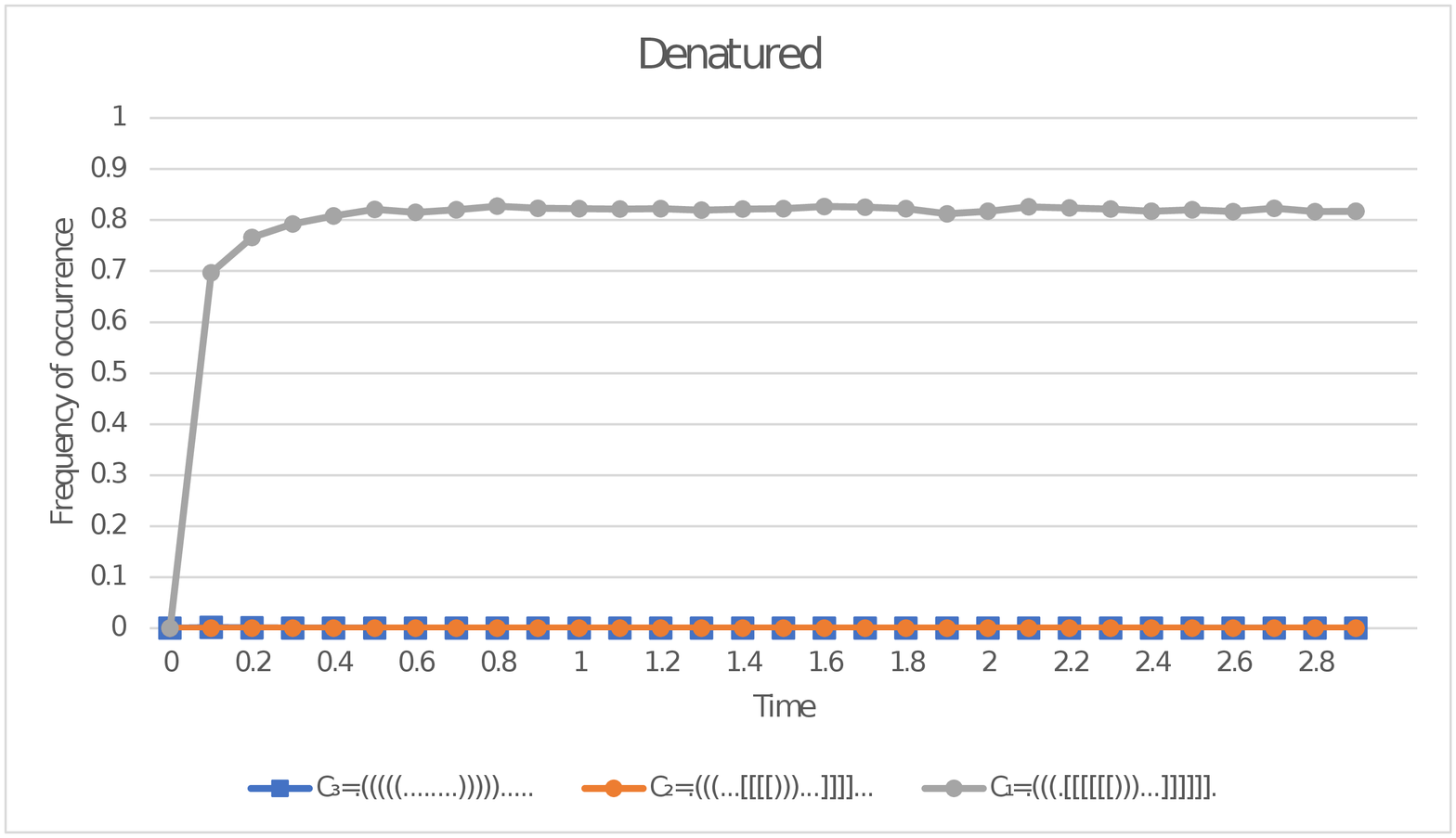}
	\caption[Folding  of the Beet soil-borne virus from denatured state]{Folding of the Beet soil-borne virus from the denatured initial state. Time on the x-axis is in seconds of simulated time.}
	\label{fig:DenaturedPseudo}       
\end{figure}

\highlightred{We use the beet soil-borne virus S = "CGGUAGCGCGAACCGUUAUCGCGCA" from the PseudoBase++ database~\cite{pseudobase++} to demonstrate the application of our simulation in predicting RNA structures with pseudoknots. The folding of the sequence S was simulated with $10000$ runs. We evaluate the energy of pseudoknots using the energy parameters from~\cite{Andronescu_2010}, estimated by fitting the standard NNDB parameters by~\cite{Mathews_1999} and pseudoknotted parameters by~\cite{Dirks_2003} over a large data set of both pseudoknotted and pseudoknot-free secondary structures.} We compare two simulation settings: a) cotranscriptional folding of S with transcription speed $200$ nt/sec, and b) the folding starting from the denatured initial state \highlightred{(i.e., a fully synthesized open strand)}. \highlightred{Figs.~\ref{fig:CotransPseudo} -~\ref{fig:DenaturedPseudo} depict the occurrence frequency of the H-type pseudoknotted structure $C_1 = $".(((.[[[[[[)))...]]]]]]." with an energy of $-12.39$ (kcal). We also consider two intermediate structures $C_2 = $=".(((...[[[[)))...]]]]..." and $C_3 = $".(((((........)))))....." having energies of $-7.25$ (kcal) and $-4.52$ (kcal), respectively.}

\highlightred{Figs.~\ref{fig:CotransPseudo} -~\ref{fig:DenaturedPseudo} clearly show that the dominant structure of the beet soil-borne virus sequence S is the H-type pseudoknotted structure $C_1$. We also see from these figures that the folding starting from the denatured state misses the formation of intermediate structures $C_2$ and $C_3$, which appear in the cotranscriptional folding. After the transcription phase, intermediate structures will rearrange to $C_1$ and remain in this stable form. Figs.~\ref{fig:CotransPseudo} shows that the frequency of $C_1$ is more than $82\%$ in the simulation.}

\highlightred{We conclude this section with a note about the energy parameters for RNA structures with pseudoknots. In particular, we also simulated the beet soil-borne virus S with the energy model by~\cite{Reidys_2011}, which is another an extension of the NNDB model for pseudoknots. The occurrence frequency of pseudoknotted structure $C_1$ estimated by the~\cite{Reidys_2011} model was significantly lower than by the~\cite{Andronescu_2010} model. This prediction discrepancy is because the energy model by~\cite{Reidys_2011} penalizes the formation of pseudoknots significantly more than the model by~\cite{Andronescu_2010}. In fact, all pseudoknotted structures will be unfavourable with such high penalties for the pseudoknots. An interesting prediction from our cotranscriptional folding simulation using both energy models is the occurrence of the intermediate hairpin structure $C_3$. The persistence of $C_3$ before rearranging to the pseudoknot $C_1$ depends on how much penalty is applied to the formation of pseudoknots.}  

\subsection{SV-11}
\begin{figure*}[!htbp]
	\centering
	\includegraphics[scale = 1.0]{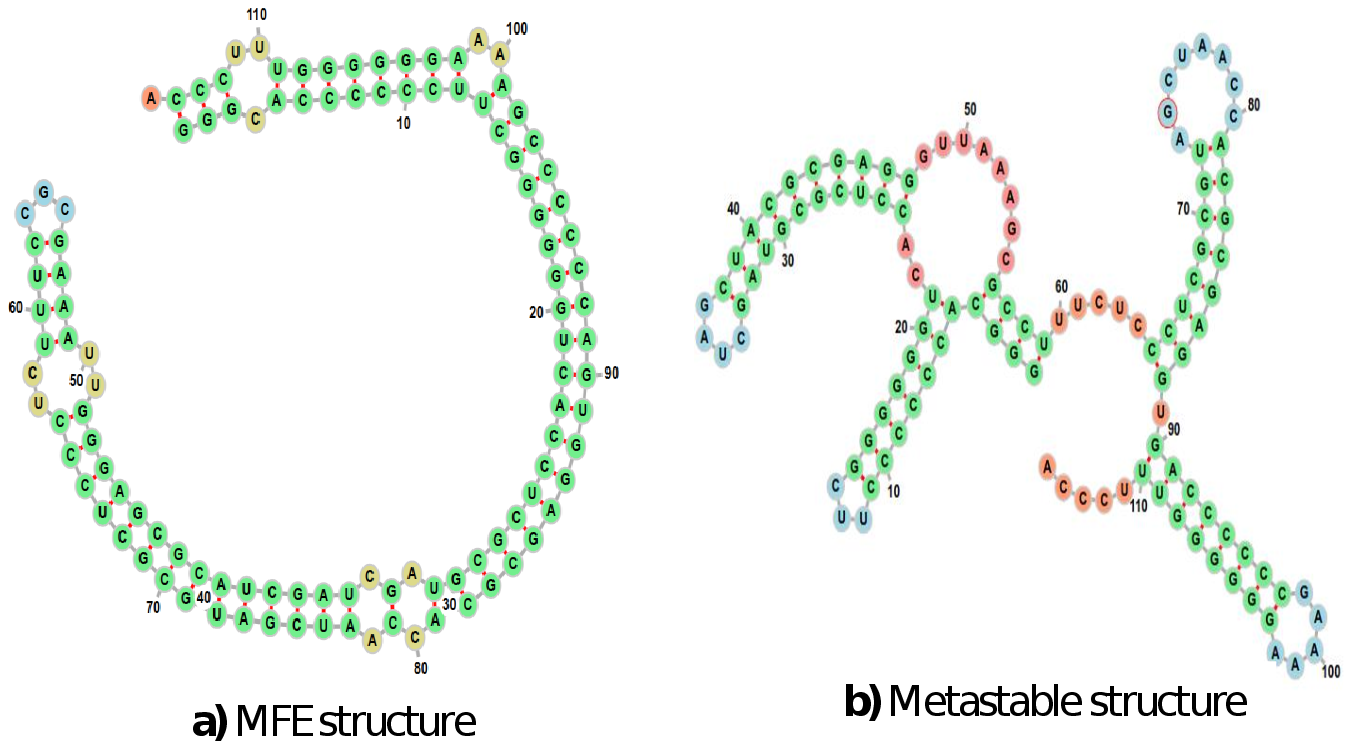}
	\caption[SV-11 with two conformations]{SV-11 with two conformations a) MFE structure ($-95.90$ kcal) and b) metastable structure ($-63.60$ kcal). The visualization of structures is made by the Forna tool~\cite{Kerpedjiev_2015}.}
	\label{fig:QBeta-1}       
\end{figure*}

SV-11 is a $115$ nt long RNA sequence. It is a recombinant between the plus and minus strands of the natural Q$\beta$ template MNV-11 RNA~\cite{Biebricher_1992}. The result of the recombination is a highly palindromic sequence whose most stable secondary structure is a long hairpin-like structure, the MFE structure in Fig.~\ref{fig:QBeta-1}a). The MFE structure, however, disables Q$\beta$ replicase because its primer regions are blocked. Experimental work~\cite{Biebricher_1992} has shown that an active structure of SV-11 for replication is when it folds into a metastable conformation depicted in Fig.~\ref{fig:QBeta-1}b). This is a hairpin-hairpin-multi-loop motif with open primer regions that serve as templates for replication. Transition from the metastable structure to the MFE structure has been observed experimentally but is rather slow~\cite{Biebricher_1992}, indicating long relaxation time to equilibrium.

\begin{figure*}[!htbp]
	\centering
	\includegraphics[scale = 0.85]{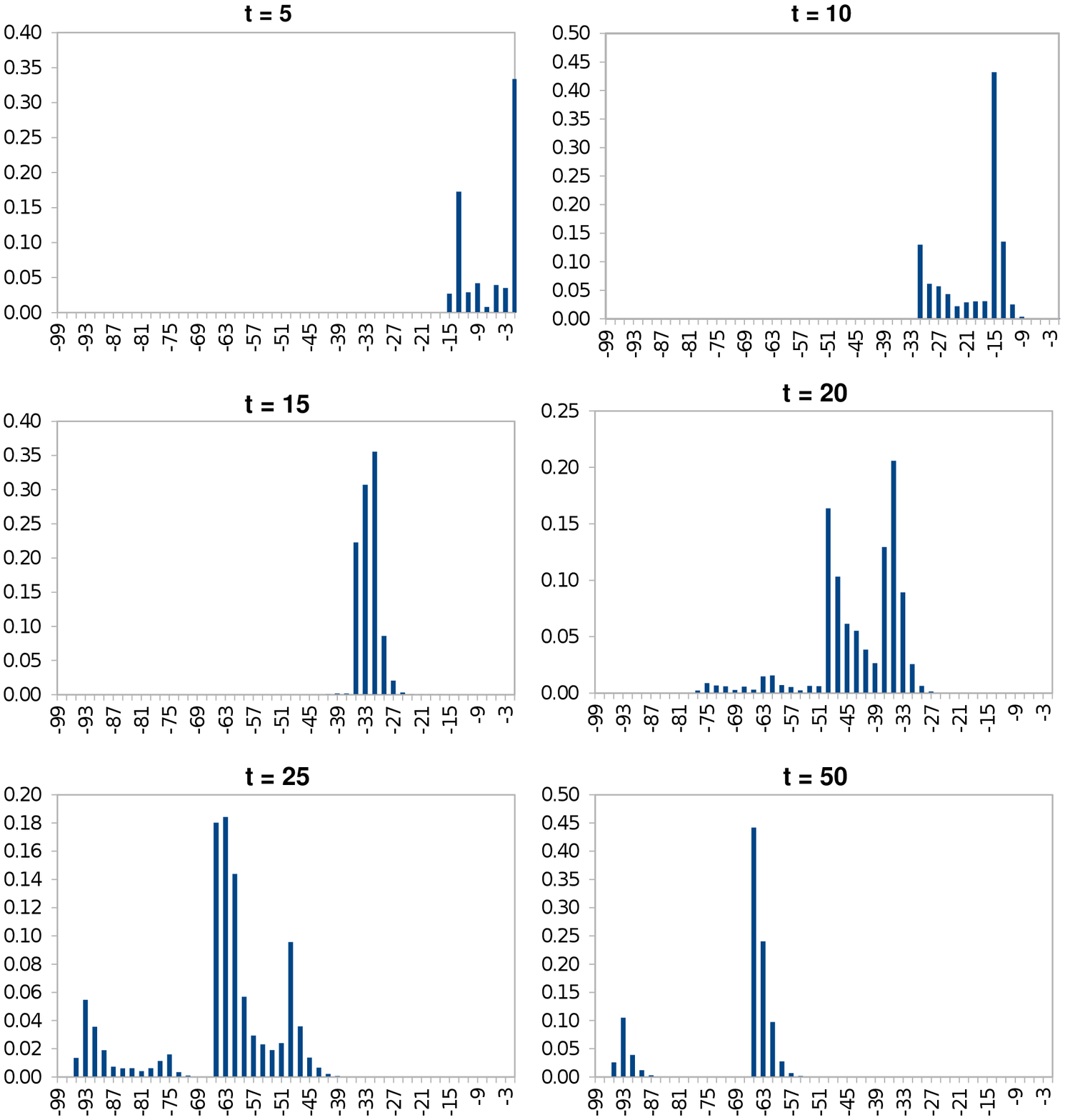}
	\caption[Cotranscriptional folding of SV-11]{Cotranscriptional folding of SV-11. The x-axis denotes the energy level in kcal, and y-axis shows the frequency of structures at a given energy level.}
	\label{fig:energy-frequency_Qbeta-1}       
\end{figure*}

We plot in Fig.~\ref{fig:energy-frequency_Qbeta-1} the energy vs. occurrence frequency of structures by the cotranscriptional folding of SV-11. The result is obtained by $10000$ simulation runs of our CoStochFold algorithm for $t = 50$s simulated seconds and average transcription speed $5$nt/sec. To determine the frequency of occurrence of a structure, we discretize the simulation time into intervals and record how much time was spent in each structure within each interval. The frequency of occurrence of a structure in each time interval is then averaged over $10000$ runs. The figure shows that the folding favors metastable structures, and disfavors the MFE structure. In particular, cotranscriptional folding quickly folds SV-11 to its metastable conformations with the mode of the energy distribution at about $-63$kcal. 

\begin{figure*}[!htbp]
	\centering 
	\includegraphics[scale = 1.0]{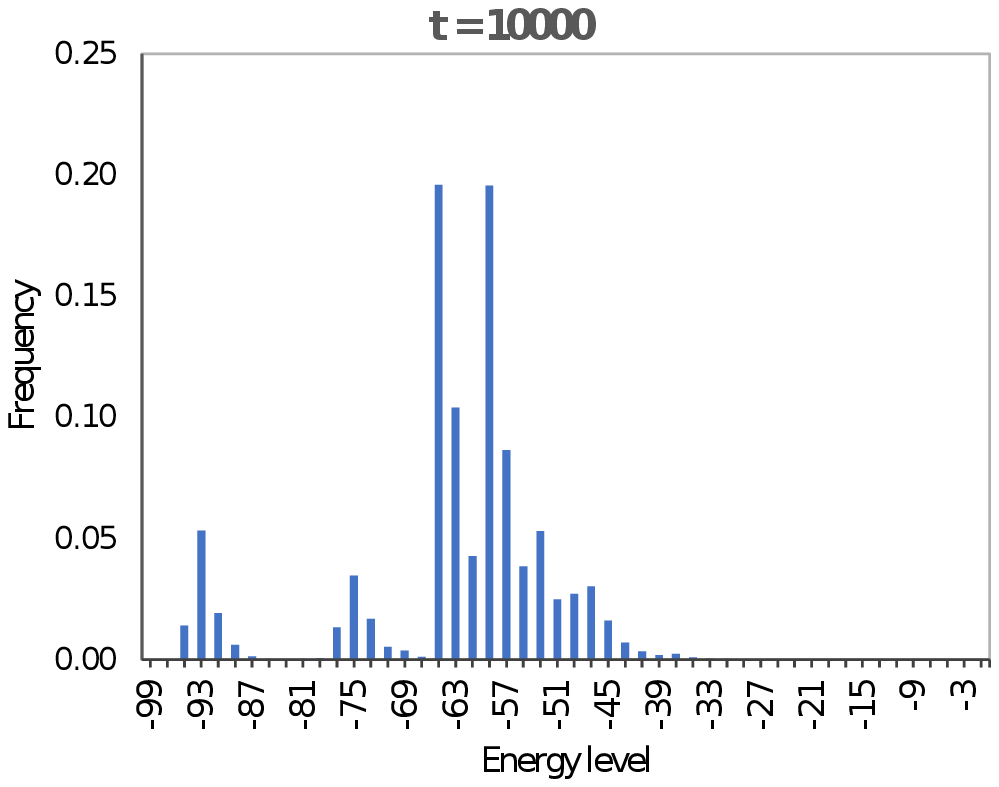}
	\caption{Frequency of structures in folding SV-11.}
	\label{fig:energy-frequency_Qbeta-2}       
\end{figure*}

\begin{figure*}[!htbp]
	\centering
	\includegraphics[scale = 1.0]{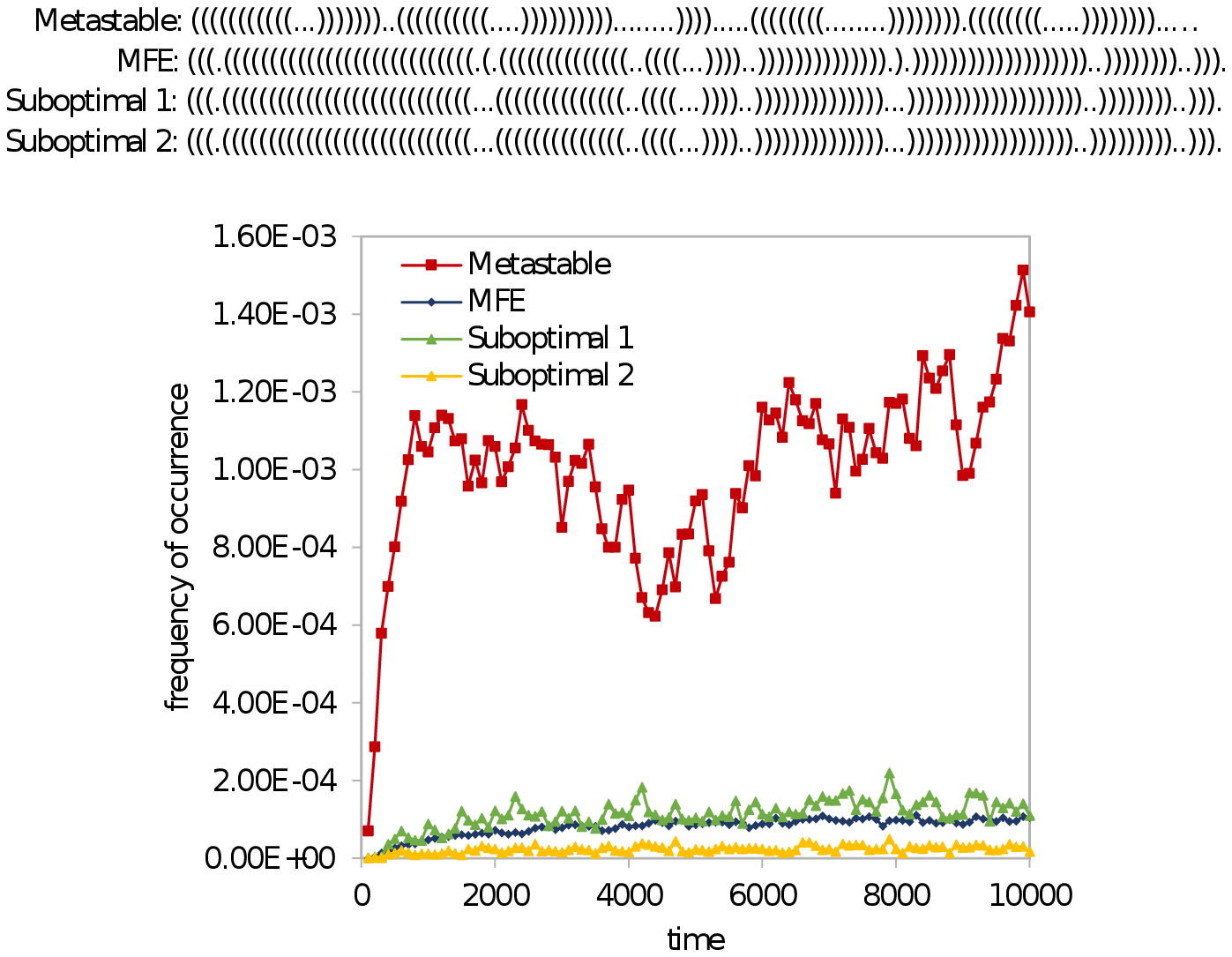}
	\caption[Frequency of the metastable structure in comparison with the MFE structure]{Frequency of the metastable structure in comparison with the MFE structure and two randomly selected suboptimal structures in the locality of the energy level of MFE.}
	\label{fig:energy-frequency_Qbeta-3}       
\end{figure*}

Fig.~\ref{fig:energy-frequency_Qbeta-2} shows the long-term occurrence frequencies of structures at different energy levels in the SV-11 folding and Fig.~\ref{fig:energy-frequency_Qbeta-3} compares the occurrence frequencies of the specific metastable structure depicted in Fig.~\ref{fig:QBeta-1}b) with the MFE structure and two randomly selected suboptimal structures in the energy level of MFE structure. Fig.~\ref{fig:energy-frequency_Qbeta-3} shows that the SV-11 molecule interestingly prefers the metastable structure over the MFE structure. Specifically, the metastable structure in the cotranscriptional folding regime is in the time interval $[0, 10000]$ about tenfold more frequent than the MFE structure.

\highlightred{
\subsection{Simulation performance}
This section reports the performance of our stochastic folding algorithm with RNA sequences of varying lengths from $25$ to $5000$nt. For each length, we executed $10$ simulation runs, each with a random sequence of the given length. We performed $1000$ simulation steps for each simulation run. The simulation was run on an Intel i5 processor. The average computational runtime for each sequence length was computed, and then divided by the number of simulation steps to normalize the computation cost.}  

\begin{figure*}[!htbp]
	\centering 
	\includegraphics[scale = 1.0]{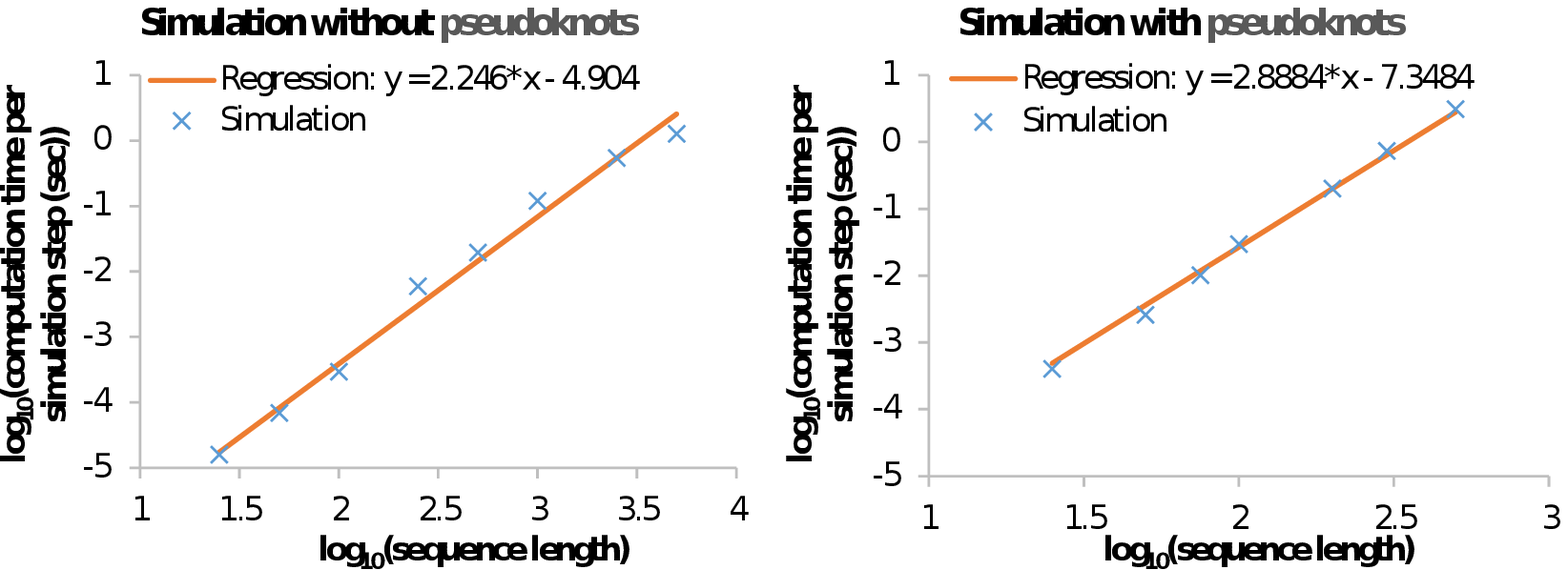}
	\caption[Computational runtimes of stochastic folding with sequences of varying lengths]{Computational runtimes of stochastic folding with sequences of varying lengths. Left: simulation without pseudoknots. Right: simulation with pseudoknots. Values in x-axis and y-axis are in log scale.}
	\label{fig:sim-performance}       
\end{figure*}

\highlightred{Fig.~\ref{fig:sim-performance} plots computational runtimes of our folding algorithm in two settings: 1) simulation without pseudoknots on the left and 2) simulation with pseudoknots on the right. As shown in the figure, the computational cost is often very computational intensive, especially for long sequences. For example, the simulation without pseudoknots for the sequence of length $1000$nt took on average $0.1$ second to execute one simulation step. A simulation experiment on our machine for the sequence of length $1000$nt with $10000$ replicas, each with $1000$ simulation steps, will thus take about $11$ days to complete. Moreover, the computational cost dramatically increases when simulating long sequences. The computational runtime in case of pseudoknot-free simulation for sequences of length $5000$ is about $11$ times higher than for sequences of length $1000$. Such intensive computational cost is due to the huge number of possible moves in the locality of a conformation. Our detailed analysis of the computational runtime shows that the cost for enumerating the possible moves contributes more than $95\%$ of the total computational cost in each simulation step. Theoretically, the number of possible moves in the locality of a conformation increases quadratically with the sequence length, i.e.,~$O(N^2)$ where $N$ is sequence length. The regression lines in Fig.~\ref{fig:sim-performance} confirmed that the computational runtimes of our algorithm is asymptotic to the theoretical result.}

\section{Conclusions}
\label{sec:5}
We propose a kinetic model of RNA folding that takes into account the elongation of an RNA chain during transcription as a primitive structure-forming operation alongside the common base-pairing operations. We developed a new stochastic simulation algorithm CoStochFold to explore RNA structure formation, including pseudoknots, in the cotranscriptional folding regime. We showed through numerical case studies that our method can quantitatively predict the formation of (metastable) conformations in an RNA folding pathway. \highlightred{The simulation method thus promises to offer useful insights into RNA folding kinetics in real biological systems. However, it also poses a great computational challenge for long sequences due to the huge number of possible moves in the locality of a conformation. Furthermore, many simulation runs must be performed in order to obtain a reasonable statistical estimation of the system dynamics. Several improvements are possible in future work. For instance, we can reduce the enumeration of possible moves by localizing the computation. The motif tree, a coarse-grained representation for pseudoknotted structures developed in the paper, could be useful also in this context. We decompose an RNA structure into motifs and then enumerate new conformations related to each motif. To reduce the cost for executing many simulation runs, we can employ high performance computing to run simulations in parallel.}   

\section*{Acknowledgements}
This work has been supported by Academy of Finland grant no. 311639, "Algorithmic Designs for Biomolecular Nano\-structures (ALBION)". The work of VHT has been partially done when he was at Aalto University. 

\bibliography{ref}

\begin{thebibliography}{45}
\providecommand{\natexlab}[1]{#1}
\providecommand{\url}[1]{\texttt{#1}}
\providecommand{\urlprefix}{URL }

\bibitem[{Akutsu(2000)}]{Akutsu_2000}
Akutsu, T., 2000.
\newblock Dynamic programming algorithms for rna secondary structure prediction
  with pseudoknots.
\newblock \emph{Discrete Appl. Math.} 104, 45--62.

\bibitem[{Al-Hashimi and Walter(2008)}]{Hashimi_2008}
Al-Hashimi, H.~M. and Walter, N.~G., 2008.
\newblock {RNA} dynamics: {It} is about time.
\newblock \emph{Curr. Opin. Struct. Biol.} 18, 321--329.

\bibitem[{Andronescu \emph{et~al.}(2010)Andronescu, Pop, and
  Condon}]{Andronescu_2010}
Andronescu, M.~S., Pop, C., and Condon, A.~E., 2010.
\newblock Improved free energy parameters for {RNA} pseudoknotted secondary
  structure prediction.
\newblock \emph{RNA} 16, 26--42.

\bibitem[{Biebricher and Luce(1992)}]{Biebricher_1992}
Biebricher, C.~K. and Luce, R., 1992.
\newblock In vitro recombination and terminal elongation of {RNA} by {Q$\beta$}
  replicase.
\newblock \emph{EMBO J.} 11, 5129--5135.

\bibitem[{Bratsun \emph{et~al.}(2005)Bratsun, Volfson, Tsimring, and
  Hasty}]{Bratsun_2005}
Bratsun, D., Volfson, D., Tsimring, L.~S., and Hasty, J., 2005.
\newblock Delay-induced stochastic oscillations in gene regulation.
\newblock \emph{PNAS} 102, 14593--14598.

\bibitem[{Chen \emph{et~al.}(2009)Chen, Condon, and Jabbari}]{Chen_2009}
Chen, H.-L., Condon, A., and Jabbari, H., 2009.
\newblock An {$O(n^5)$} algorithm for {MFE} prediction of kissing hairpins and
  4-chains in nucleic acids.
\newblock \emph{J. Comp. Biol.} 16, 803--815.

\bibitem[{Collins and Penny(2009)}]{Collins_2009}
Collins, L.~J. and Penny, D., 2009.
\newblock The {RNA} infrastructure: Dark matter of the eukaryotic cell?
\newblock \emph{Trends Genet.} 25, 120--128.

\bibitem[{Dill(1999)}]{Dill_1999}
Dill, K.~A., 1999.
\newblock Polymer principles and protein folding.
\newblock \emph{Protein Science} 8, 1166--1180.

\bibitem[{Dirks and Pierce(2003)}]{Dirks_2003}
Dirks, R.~M. and Pierce, N.~A., 2003.
\newblock A partition function algorithm for nucleic acid secondary structure
  including pseudoknots.
\newblock \emph{J. Comp. Chem.} 24, 1664--1677.

\bibitem[{Dykeman(2015)}]{Dykeman_2015}
Dykeman, E.~C., 2015.
\newblock An implementation of the {Gillespie} algorithm for {RNA} kinetics
  with logarithmic time update.
\newblock \emph{Nucleic Acids Res.} 43, 5708--5715.

\bibitem[{Fallmann \emph{et~al.}(2017)Fallmann, Will, Engelhardt, Gr\"{u}ning,
  Backofenc, and Stadler}]{Fallmann_2017}
Fallmann, J., Will, S., Engelhardt, J., Gr\"{u}ning, B., Backofenc, R., and
  Stadler, P.~F., 2017.
\newblock Recent advances in {RNA} folding.
\newblock \emph{J. Biotechnol.} 261, 97--104.

\bibitem[{Flamm \emph{et~al.}(2000)Flamm, Fontana, Hofacker, and
  Schuster}]{Flamm_2000}
Flamm, C., Fontana, W., Hofacker, I.~L., and Schuster, P., 2000.
\newblock {RNA} folding at elementary step resolution.
\newblock \emph{RNA} 6, 325--338.

\bibitem[{Gultyaev \emph{et~al.}(1995)Gultyaev, van Batenburg F. H.~D., and
  Pleij}]{Gultyaev_1995}
Gultyaev, A.~P., van Batenburg F. H.~D., and Pleij, C. W.~A., 1995.
\newblock The computer simulation of {RNA} folding pathways using a genetic
  algorithm.
\newblock \emph{J. Mol. Biol.} 250, 37--51.

\bibitem[{Hofacker \emph{et~al.}(1998)Hofacker, Schuster, and
  Stadler}]{Hofacker_1998}
Hofacker, I.~L., Schuster, P., and Stadler, P.~F., 1998.
\newblock Combinatorics of {RNA} secondary structures.
\newblock \emph{Discrete Appl. Math.} 88, 207--237.

\bibitem[{Hofacker and Stadler(2005)}]{Hofacker_2005}
Hofacker, I.~L. and Stadler, P.~F., 2005.
\newblock {RNA} secondary structures.
\newblock In Meyers, R.~A., ed., \emph{Encyclopedia of Molecular Cell Biology
  and Molecular Medicine, Volume 12}, 581--603. Wiley-VCH Verlag GmbH.

\bibitem[{Hua \emph{et~al.}(2018)Hua, Panja, Wang, Woodson, and Ha}]{Hua_2018}
Hua, B., Panja, S., Wang, Y., Woodson, S.~A., and Ha, T., 2018.
\newblock Mimicking co-transcriptional rna folding using a superhelicase.
\newblock \emph{J. Am. Chem. Soc.} 140, 10067--10070.

\bibitem[{Isambert and Siggia(2000)}]{Isambert_2000}
Isambert, H. and Siggia, E.~D., 2000.
\newblock Modeling {RNA} folding paths with pseudoknots: {Application} to
  hepatitis delta virus ribozyme.
\newblock \emph{PNAS} 97, 6515.

\bibitem[{Jasinski \emph{et~al.}(2017)Jasinski, Haque, Binzel, and
  Guo}]{Jasinski_2017}
Jasinski, D., Haque, F., Binzel, D.~W., and Guo, P., 2017.
\newblock Advancement of the emerging field of {RNA} nanotechnology.
\newblock \emph{ACS Nano} 11, 1142--1164.

\bibitem[{Kawasaki(1966)}]{Kawasaki_1966}
Kawasaki, K., 1966.
\newblock Diffusion constants near the critical point for time-dependent
  {Ising} models.
\newblock \emph{Phys. Rev.} 145, 224--230.

\bibitem[{Kerpedjiev \emph{et~al.}(2015)Kerpedjiev, Hammer, and
  Hofacker}]{Kerpedjiev_2015}
Kerpedjiev, P., Hammer, S., and Hofacker, I.~L., 2015.
\newblock Forna (force-directed {RNA}): Simple and effective online {RNA}
  secondary structure diagrams.
\newblock \emph{Bioinformatics} 31, 3377--3379.

\bibitem[{Lai \emph{et~al.}(2013)Lai, Proctor, and Meyer}]{Lai_2013}
Lai, D., Proctor, J.~R., and Meyer, I.~M., 2013.
\newblock On the importance of cotranscriptional {RNA} structure formation.
\newblock \emph{RNA} 19, 1461--1473.

\bibitem[{Lyngs\o\ and Pedersen(2000)}]{Lyngso_2000}
Lyngs\o\, R.~B. and Pedersen, C. N.~S., 2000.
\newblock {RNA} pseudoknot prediction in energy-based models.
\newblock \emph{J. Comp. Biol.} 7, 409--427.

\bibitem[{Marchetti \emph{et~al.}(2016)Marchetti, Priami, and
  Thanh}]{Marchetti_2016}
Marchetti, L., Priami, C., and Thanh, V.~H., 2016.
\newblock {HRSSA}–efficient hybrid stochastic simulation for spatially
  homogeneous biochemical reaction networks.
\newblock \emph{J. Comp. Phys.} 317, 301--317.

\bibitem[{Marchetti \emph{et~al.}(2017)Marchetti, Priami, and
  Thanh}]{Marchetti_2017}
Marchetti, L., Priami, C., and Thanh, V.~H., 2017.
\newblock \emph{Simulation Algorithms for Computational Systems Biology}.
\newblock Springer.

\bibitem[{Mathews \emph{et~al.}(1999)Mathews, Sabina, Zuker, and
  Turner}]{Mathews_1999}
Mathews, D.~H., Sabina, J., Zuker, M., and Turner, D.~H., 1999.
\newblock Expanded sequence dependence of thermodynamic parameters improves
  prediction of {RNA} secondary structure.
\newblock \emph{J. Mol. Biol.} 288, 911--940.

\bibitem[{Metropolis \emph{et~al.}(1953)Metropolis, Rosenbluth, Rosenbluth, and
  Teller}]{Metropolis_1953}
Metropolis, N., Rosenbluth, A.~W., Rosenbluth, M.~N., and Teller, A.~H., 1953.
\newblock Equation of state calculations by fast computing machines.
\newblock \emph{J. Chem. Phys.} 21, 1087--1092.

\bibitem[{Mironov and Kister(1986)}]{Mironov_1986}
Mironov, A. and Kister, A., 1986.
\newblock {RNA} secondary structure formation during transcription.
\newblock \emph{Journal of Biomolecular Structure and Dynamics} 4, 1--9.

\bibitem[{Mironov and Lebedev(1993)}]{Mironov_1993}
Mironov, A.~A. and Lebedev, V.~F., 1993.
\newblock A kinetic model of {RNA} folding.
\newblock \emph{Biosystems} 30, 49--56.

\bibitem[{Pan and Sosnick(2006)}]{Pan_2006}
Pan, T. and Sosnick, T.~R., 2006.
\newblock {RNA} folding during transcription.
\newblock \emph{Annu. Rev. Biophys. Biomol. Struct.} 35, 161--175.

\bibitem[{Proctor and Meyer(2013)}]{Proctor_2013}
Proctor, J.~R. and Meyer, I.~M., 2013.
\newblock {COFOLD}: an {RNA} secondary structure prediction method that takes
  co-transcriptional folding into account.
\newblock \emph{Nucleic Acids Res.} 41, e102.

\bibitem[{Rastegari and Condon(2007)}]{Condon_2007}
Rastegari, B. and Condon, A., 2007.
\newblock Parsing nucleic acid pseudoknotted secondary structure: Algorithm and
  applications.
\newblock \emph{Journal of Computational Biology} 14, 16--32.

\bibitem[{Reeder and Giegerich(2004)}]{Reeder_2004}
Reeder, J. and Giegerich, R., 2004.
\newblock Design, implementation and evaluation of a practical pseudoknot
  folding algorithm based on thermodynamics.
\newblock \emph{BMC Bioinformatics} 5.

\bibitem[{Reidys \emph{et~al.}(2011)Reidys, Huang, Andersen, Penner, Stadler,
  and Nebel}]{Reidys_2011}
Reidys, C.~M., Huang, F. W.~D., Andersen, J.~E., Penner, R.~C., Stadler, P.~F.,
  and Nebel, M.~E., 2011.
\newblock Topology and prediction of {RNA} pseudoknots.
\newblock \emph{Bioinformatics} 27, 1076--1085.

\bibitem[{Rivas and Eddy(1999)}]{Rivas_1999}
Rivas, E. and Eddy, S.~R., 1999.
\newblock A dynamic programming algorithm for {RNA} structure prediction
  including pseudoknots.
\newblock \emph{J. Mol. Biol.} 285, 2053--2068.

\bibitem[{Storz(2002)}]{Storz_2002}
Storz, G., 2002.
\newblock An expanding universe of noncoding {RNA}s.
\newblock \emph{Science} 296, 1260--1263.

\bibitem[{Taufer \emph{et~al.}(2008)Taufer, Licon, Araiza, Mireles, van
  Batenburg, Gultyaev, and Leung}]{pseudobase++}
Taufer, M., Licon, A., Araiza, R., Mireles, D., van Batenburg, F. H.~D.,
  Gultyaev, A.~P., and Leung, M.-Y., 2008.
\newblock {PseudoBase++}: an extension of {PseudoBase} for easy searching,
  formatting and visualization of pseudoknots.
\newblock \emph{Nucleic Acids Research} 37, D127--D135.

\bibitem[{Thanh \emph{et~al.}(2014)Thanh, Priami, and Zunino}]{Thanh_2014_2}
Thanh, V.~H., Priami, C., and Zunino, R., 2014.
\newblock Efficient rejection-based simulation of biochemical reactions with
  stochastic noise and delays.
\newblock \emph{J. Chem. Phys.} 141, 10B602.

\bibitem[{Thanh and Zunino(2014)}]{Thanh_2014}
Thanh, V.~H. and Zunino, R., 2014.
\newblock Adaptive tree-based search for stochastic simulation algorithm.
\newblock \emph{Int. J. Comput. Biol. Drug. Des.} 74, 341--357.

\bibitem[{Thanh \emph{et~al.}(2016)Thanh, Zunino, and Priami}]{Thanh_2016}
Thanh, V.~H., Zunino, R., and Priami, C., 2016.
\newblock Efficient constant-time complexity algorithm for stochastic
  simulation of large reaction networks.
\newblock \emph{IEEE/ACM Trans. Comput. Biol. Bioinform.} 14, 657--667.

\bibitem[{Thanh \emph{et~al.}(2017)Thanh, Zunino, and Priami}]{Thanh_2017}
Thanh, V.~H., Zunino, R., and Priami, C., 2017.
\newblock Efficient stochastic simulation of biochemical reactions with noise
  and delays.
\newblock \emph{J. Chem. Phys.} 146, 084107.

\bibitem[{Turner and Mathews(2009)}]{Turner_2009}
Turner, D.~H. and Mathews, D.~H., 2009.
\newblock {NNDB}: The nearest neighbor parameter database for predicting
  stability of nucleic acid secondary structure.
\newblock \emph{Nucleic Acids Res.} 38, D280--D282.

\bibitem[{Watters \emph{et~al.}(2016)Watters, Strobel, Yu, Lis, and
  Lucks}]{Watters_2016}
Watters, K.~E., Strobel, E.~J., Yu, A.~M., Lis, J.~T., and Lucks, J.~B., 2016.
\newblock Cotranscriptional folding of a riboswitch at nucleotide resolution.
\newblock \emph{Nat. Struct. Mol. Biol.} 23, 1124--1131.

\bibitem[{Zhao \emph{et~al.}(2011)Zhao, Zhang, and Chen}]{Zhao_2011}
Zhao, P., Zhang, W., and Chen, S.-J., 2011.
\newblock Cotranscriptional folding kinetics of ribonucleic acid secondary
  structure.
\newblock \emph{J. Chem. Phys} 135, 245101.

\bibitem[{Zuker(1989)}]{Zuker_1989}
Zuker, M., 1989.
\newblock On finding all suboptimal foldings of an {RNA} molecule.
\newblock \emph{Science} 244, 48--52.

\bibitem[{Zuker and Stiegler(1981)}]{Zuker_1981}
Zuker, M. and Stiegler, P., 1981.
\newblock Optimal computer folding of large {RNA} sequences using
  thermodynamics and auxiliary information.
\newblock \emph{Nucleic Acids Res.} 9, 133--148.

\end{thebibliography}

\end{document}